\documentclass[onecolumn,amsmath,showpacs,nofootinbib,11pt]{revtex4}
\usepackage{mathrsfs} 
\usepackage{graphicx}
\usepackage{bm}
\usepackage{floatrow}
\usepackage{fancyhdr}
\usepackage{longtable}
\usepackage{hyperref}
\usepackage{subfigure}
\usepackage{epsfig} 
\usepackage{float}
\usepackage{color,soul}
\usepackage{menukeys}
\usepackage{xcolor}
\usepackage{feynman}
\usepackage{amssymb}
\usepackage{braket}
\usepackage{graphicx}
\usepackage{dcolumn}
\usepackage[mathscr]{eucal}
\usepackage{amsmath}
\usepackage{tabularx}
\usepackage{booktabs}
\usepackage{slashed}
\usepackage{mathtools,slashed}
\usepackage[utf8x]{inputenc}

\makeatletter
\def\pslashed#1{%
\expandafter\ifx\csname psla@\string#1\endcsname\relax
{\mathpalette{\sla@/00}{\phantom{#1}}}%
\else
\csname psla@\string#1\endcsname\fi}
\def\declarepslashed#1#2#3#4#5{%
\expandafter\def\csname psla@\string#5\endcsname{%
#1{\mathpalette{\sla@{#2}{#3}{#4}}{\phantom{#5}}}}}
\makeatother
\declarepslashed{}{/}{.08}{0}{D}

\newcommand{\hs}{\hspace*{0.5cm}}

\newcommand{\be}{\begin{equation}}
	\newcommand{\ee}{\end{equation}}
\newcommand{\bea}{\begin{eqnarray}}
	\newcommand{\eea}{\end{eqnarray}}
\newcommand{\ben}{\begin{enumerate}}
	\newcommand{\een}{\end{enumerate}}
\newcommand{\bde}{\begin{widetext}}
	\newcommand{\ede}{\end{widetext}}
\newcommand{\nn}{\nonumber}
\newcommand{\crn}{\nonumber \\}

\newcommand{\al}{\alpha}

\newcommand{\fr}{\frac}
\newcommand{\bc}{\begin{center}}
	\newcommand{\ec}{\end{center}}
\newcommand{\Ga}{\Gamma}
\newcommand{\ga}{\gamma}

\newcommand{\ep}{\epsilon}

\newcommand{\La}{\Lambda}

\begin{document}

\title{ Lepton universality violation in the MF331 model}
\author{P. N. Thu $^{a,c}$}
\author{N. T. Duy $^{a,b}$}
\author{A. E. C\'arcamo Hern\'andez$^{d,e,f}$}
\author{D. T. Huong$^{b}$}
\email{dthuong@iop.vast.vn (Corresponding author)}
\email{thupn@utb.edu.vn}
\email{ntduy@iop.vast.vn}
\email{antonio.carcamo@usm.cl}
\affiliation{$^a$ Graduate University of Science and Technology,
	Vietnam Academy of Science and Technology,
	18 Hoang Quoc Viet, Cau Giay, Hanoi, Vietnam \\$^b$ Institute of Physics, VAST, 10 Dao Tan, Ba Dinh, Hanoi, Vietnam\\
	$^c$ Faculty of Natural Sciences and Technology
	Tay Bac University, Quyet Tam Ward, Son La City, Son La Province \\
 $^d$ Universidad T\'{e}cnica Federico Santa Mar\'{\i}a, Casilla 110-V,
 Valpara\'{\i}so, Chile\\
$^e$ Centro Cient\'{\i}fico-Tecnol\'ogico de Valpara\'{\i}so, Casilla 110-V, Valpara\'{\i}so, Chile\\
$^f$ Millennium Institute for Subatomic physics at high energy frontier - SAPHIR, Fernandez Concha 700, Santiago, Chile}
\date{\today}

\date{\today}

\begin{abstract}
We perform a detailed study of the $\text{b} \to \text{c} \tau \nu$ and $\text{b} \to \text{s} l^+ l^-$ processes in a minimal flipped 331 model based on the $SU(3)_C\times SU(3)_L\times U(1)_N$ gauge symmetry. The non universal $SU(3)_L \times U(1)_N$ symmetry in the lepton sector gives rise to non universal neutral and charged currents involving heavy non SM gauge bosons and SM leptons that yield radiative contributions to the  $b \to s$, $b\to c$, $s\to u$ and $d\to u$ transitions arising from one loop level penguin and box diagrams. We found that the observables related to these transitions agree with their experimental values in a region of parameter space that includes TeV scale exotic up type quarks, within the LHC's reach.
\end{abstract}

\maketitle
	\section{Introduction} 
In recent years, experimental data in B physics has hinted toward deviations of Lepton Flavor Universality (LFU) in semi-leptonic decays from the Standard Model (SM) expectations. More specifically, measurements of $\text{V}_{\text{cb}}-$independent ratios
\bea
\text{R}(\text{D}^{(*)})= \frac{\mathcal{B}(\text{B} \to \text{D}^{(*)} \tau \nu)}
{\mathcal{B}(\text{B} \to \text{D}^{(*)} l \nu)},
\eea 
with $l=e$ or $\mu$, have been performed by the Babar \cite{BaBar:2012obs,BaBar:2013mob}, Belle \cite{Belle:2015qfa,Belle:2016kgw,Abdesselam:2016xqt}, and LHCb \cite{LHCb:2015gmp} collaborations. The world average result, which is extracted from the latest announcement of LHCb, is given as:
\bea
\text{R}(\text{D})_{\text{exp}} && =0.356 \pm 0.029_{\text{total}}  \hs
\text{R}(\text{D}^{(*)})_{\text{exp}} =0.284 \pm 0.013_{\text{total}},
\label{tnph1}\eea
On the other hand, the SM calculations for these ratios, which are performed by several groups \cite{Bigi:2016mdz,Fajfer:2012vx,Becirevic:2012jf,Bernlochner:2017jka,Bigi:2017jbd,Jaiswal:2017rve}, are averaged  by \cite{HFLAV:2016hnz} 
\bea
\text{R}(\text{D})_{\text{SM}} && =0.298\pm 0.004, \hs 
\text{R}(\text{D}^{(*)})_{\text{SM}}  =0.254  \pm 0.005.
\label{RDSM}\eea
These relative rates have been predicted with rather high accuracy because many hadronic uncertainties are canceled out to a large extent. The SM expectations are significantly lower than the measurements. If confirmed, this could be a signal of new physics (NP). In principle, the NP contributions could be due to a tree-level exchange of a new charged scalar \cite{Crivellin:2012ye,Celis:2012dk,Crivellin:2013wna}, a heavy charged vector \cite{Greljo:2015mma,Boucenna:2016qad,Greljo:2018ogz}, or due to an exchange of leptoquarks \cite{Dorsner:2016wpm,Bauer:2015knc,Fajfer:2015ycq,Barbieri:2015yvd,Becirevic:2016yqi,Hiller:2016kry,Crivellin:2017zlb}.
Effects due to the presence of light sterile neutrinos have
also been explored in \cite{Abada:2013aba,Cvetic:2016fbv,Crivellin:2017zlb}. Besides, the LHCb, Belle  collaborations measured \cite{Aaij_2014,Aaij_2019,Choudhury_2021,Aaij2021,Aaij_2017,Wehle_2021} 
the ratios,
\bea \text{R}_{\text{K}}&&\equiv \frac{\text{Br}\left(\text{B}^{+}\rightarrow \text{K}^{+} \mu^+ \mu^-\right)}{\text{Br}\left(\text{B}^{+}\rightarrow \text{K}^{+} e^+ e^-\right)},\hs
\text{R}_{\text{K}^{*}}  \equiv \frac{\text{Br}\left(\text{B}\rightarrow \text{K}^{*} \mu^+ \mu^-\right)}{\text{Br}\left(\text{B}\rightarrow \text{K}^{*} e^+ e^-\right)}.
\eea 
The latest values of $\text{R}_{\text{K}}, \text{R}_{\text{K}^*} $ have been reported by\cite{lhcbcollaboration2022test},\cite{lhcbcollaboration2022measurement}, such as 
\bea
\begin{cases}
	\text{R}_{\text{K}}=0.994^{+0.090}_{-0.082}    (\text{stat})^{+0.027}_{-0.029}(\text{syst})  & \text{for low-} \text{q}^2,
	\, , \\
	\text{R}_{\text{K}}=0.949^{+0.042}_{-0.041}(\text{stat})^{+0.023}_{-0.023}(\text{syst)} &  \text{for central-} \text{q}^2.
\end{cases}    
\eea
which showed $0.2 \sigma$ deviation from the SM expectation \cite{Bordone_2016, straub2018flavio} of $\simeq 1$, and 
\begin{eqnarray*}
	\text{R}_{\text{K}^{*}}^{\text{LHCb}} = 
	\begin{cases}
		0.927^{+0.093}_{-0.087}(\text{sat})^{+0.034}_{-0.033}(\text{syst})  & \text{for low-} \text{q}^2\, , \\
		1.027^{+0.072}_{-0.068}(\text{sat})^{+0.027}_{-0.027}(\text{syst}) & \text{for central-} \text{q}^2.
	\end{cases}
\end{eqnarray*} 
These ratios also are $0.2$ standard deviations from their SM expectations \cite{Bordone_2016, straub2018flavio, Altmannshofer_2017}. Solutions for both sets of anomalies are very scarce. 
This is because the semileptonic decay $\text{B} \to \text{D}^* \tau \nu$ is a charged current process that occurs at the tree level, whereas the decay process $\text{B} \to \text{K}^{(*)}ll$ occurs at the one-loop level in the SM. If the new interactions maintained the SM pattern, both processes would account for a large deviation from the SM of similar size, which again $\text{O}(25$\%$)$,  would point to a very light mediator. However, the light mediator would be hard to hide from other observables that are in perfect agreement with the SM. In fact, among the proposed models, the model based on an extended gauge symmetry group \cite{Boucenna:2016wpr}, leptoquarks \cite{Dorsner:2016wpm,Bauer:2015knc,Fajfer:2015ycq,Barbieri:2015yvd,Becirevic:2016yqi,Hiller:2016kry,Crivellin:2017zlb,Alonso:2015sja}, strong interactions \cite{Buttazzo:2016kid}, and an effective theory approach \cite{Alonso:2015sja,Bhattacharya:2014wla,Greljo:2015mma,Calibbi:2015kma} are possible to produce both solutions. Specifically,  in the non-universal gauge extensions of the SM \cite{Boucenna:2016wpr} the B-decay anomalies are explained by non-trivial tuning between parameters of the scalar and gauge sectors, and the gauge mixing terms, but the gauge mixing effects are suppressed. In this work, we show that the minimal flipped \text{3-3-1} (MF331) model \cite{Van_Loi_2020},  a version of the F331 models in which scalar multiplets are reduced to a minimum, can provide a possible explanation from SM deviations observed in B-meson decays. The F331 model is one of the extended SM  that is based on the gauge symmetry $SU(3)_C \times SU(3)_L \times U(1)_X \text{(3-3-1)}$. We call it the 331 models. The MF331 model has all the benefits of the 331 model \cite{Pisano_1992,PhysRevLett.69.2889,PhysRevD.47.4158,PhysRevD.22.738,Montero_1993,Foot_1994} due to including solutions for dark matter, neutrino masses, cosmic inflation, and matter-antimatter asymmetry, all of which are current SM issues. This also explains the existence of only three SM fermion families, strong CP conservation, and electric charge quantization. The difference between the F331 model compared to other versions of the 331 model is an arrangement of fermions in each generation. In the F331 model, the first lepton generation transforms differently from the other two lepton generations. Therefore, the model predicts non-universal interactions between the SM leptons and new particles (new fermions and new gauge bosons) \cite{Huong_2019}, which naturally provide solutions for explaining the anomalies in the $\text{B}$ meson decays. In the paper \cite{Duy:2022qhy}, we considered the NP contribution to the  $\text{R}_\text{K}$, $\text{R}_{\text{K}^*}$, and where we have looked for a parameter space that can explain anomalies \cite{Aaij2021} \cite{Aaij_2017}. We demonstrated that the anomalies $\text{R}_\text{K}$, $\text{R}_{\text{K}^*}$ given in \cite{Aaij2021} \cite{Aaij_2017} 
can be explained in the MF331 model if there is mass degeneracy between the new leptons $E_i$ and $\xi$. 
However, the new results \cite{lhcbcollaboration2022test},\cite{lhcbcollaboration2022measurement}, which have been reported at the end of 2022, show the comparison of 
$\text{R}_{\text{K}}$ and $\text{R}_{\text{K}^*}$
measured with the SM prediction. 
As a result, in this study, we revisit the NP contribution to the $\text{R}_{\text{K}}$ and $\text{R}_{\text{K}^*}$
in addition to investigating the  NP contributions of charge currents to the $\text{R}_\text{D}$, $\text{R}_{\text{D}^*}$ anomalies.
The NP effects of the charge current on the $s-u$, $d-u$ transition processes. 
We show how a simultaneous explanation of the $\text{b} \to \text{c} \tau \nu$ and $\text{b} \to \text{s} l^+ l^-$ processes can be solved in the context of the MF331 model. 
The structure of the paper is organized as follows. We perform a summary of the particle contents and their mass spectra for the  $\text{MF331}$ model in Sec. (\ref{MF331}). In Sec. \ref{current}, we examine all of the NP contributions that modify the SM-charged (neutral) currents at the tree level. Effective Hamiltonian for the flavour non-universal $u_i - d_j$ transitions are described in Sec. \ref{Eff}. In Sec.\ref{Phys}, we investigate a variety of observables related to flavour non-universality interactions. Finally, we provide our conclusions in Sec. \ref{Concl.}. 

    \section{The model\label{MF331}}
\subsection{Paticle content}
The MF331 model \cite{Fonseca_2016, Van_Loi_2020} is an unified theory with electroweak gauge group promoted to $SU(3)_L\times U(1)_N$. The electric charge and hypercharge are determined by
\bea
Q= T_3+ \frac{1}{\sqrt{3}} T_8 + X, \hs Y =\frac{1}{\sqrt{3}} T_8 + X,
\eea
with $T_3,T_8$ are the diagonal generators of $SU(3)_L$, and $X$ is a generator of $U(1)_X$. The particle content in the MF331 model is presented in Table \ref{particle}, and more details can be found in \cite{Van_Loi_2020}. 
\begin{table}[H]
	\bc
	\begin{tabular}{|c|c|c|c|}
		\hline
		\hs	Generations \hs & \hs $SU(3)_C$ \hs & \hs $ SU(3)_L$ \hs & \hs $U(1)_N$  \hs \\  
		\hline
		\hs	$\Psi_{1L}$  \hs & \hs $1$ \hs  & $6$ & $- \frac{1}{3}$ \\ \hline
		\hs	$\Psi_{\al 1L}$ \hs & \hs $2$ \hs  & $3$ & $- \frac{2}{3}$ \\ \hline
		\hs	$Q_{aL}$ \hs & \hs $3$ \hs  & $3^*$ & $ \frac{1}{3}$ \\ \hline
		\hs	$e_{aR}$ \hs & \hs $1$ \hs  & $1$ & $ -1$ \\ \hline
		\hs	$E_{aR}$ \hs & \hs $1$ \hs  & $1$ & $ -1$ \\ \hline
		\hs	$u_{aR}$ \hs & \hs $3$ \hs  & $1$ & $ \frac{2}{3}$ \\ \hline
		\hs	$d_{aR}$ \hs & \hs $3$ \hs  & $1$ & $ -\frac{1}{3}$ \\ \hline
		\hs	$U_{aR}$ \hs & \hs $3$ \hs  & $1$ & $ \frac{2}{3}$ \\ \hline
		\hs	$\rho$ \hs & \hs $1$ \hs  & $3$ & $ \frac{1}{3}$ \\ \hline
		\hs	$\chi$ \hs & \hs $1$ \hs  & $3$ & $ \frac{1}{3}$ \\ \hline
	\end{tabular}
	\caption{\label{particle} Particle content of the MF331 model where $\al=1,2$, $a=1,2,3$.}
	\ec
\end{table}

The gauge bosons and gauge couplings of the extended electroweak group are denoted as follows 
\bea
SU(3)_L && : \hs g, \hs  A_\mu^a, \hs  a=1,...,8, \nonumber \crn
U(1)_X &&: \hs g^\prime = T_{X} g, \hs B_\mu.
\eea
The model contains three neutral gauge bosons, $A_3, A_8, B$, that are mixed. By diagonalizing this mixing matrix, the model produces three physical states, such as \cite{Van_Loi_2020}
\bea A &=& s_W A_3 +  \left(\frac{c_Wt_W}{\sqrt{3}}A_8 + c_W\sqrt{1-\frac{t^2_W}{3}}B\right),\\
Z&=& c_W A_3 - \left(\frac{s_Wt_W}{\sqrt{3}}A_8 + s_W\sqrt{1-\frac{t^2_W}{3}}B\right),\\
Z'& = &\sqrt{1-\frac{t^2_W}{3}}A_8 - \fr{t_W}{\sqrt{3}}B,
\eea
and their masses are $\left( 0, \fr{g^2v^2}{4c^2_W}, \fr{g^2\left[c^2_{2W}v^2 + 4c^4_Ww^2\right]}{4c^2_W(3 - 3s^2_W)} \right)$, respectively. Due to the existence of $v', w' $, there is a slight mixing between two neutral gauge bosons, $Z, Z^\prime$, with a mixing angle defined as follows
\bea
t_{2\varphi}= -\fr{c_{2W}\sqrt{1+2c_{2W}}v^2}{2c^4_Ww^2}.
\eea 
There are six non-hermitian gauge boson states, $W^\pm= \frac{A_1\mp iA_2}{\sqrt{2}}$, $X^\pm= \frac{A_4\mp iA_5}{\sqrt{2}}, Y^{0, (0*)}= \frac{A_6 \mp i A_{7}}{\sqrt{2}}$. The presence of the VEVs, $v^\prime$, $w^\prime$, leads to the mixing of the charged gauge bosons, $W^\pm, X^\pm$. The physical states are $(W', X')$ which are determined in \cite{Van_Loi_2020} as
\bea \left\{  \begin{array}{l}
	W_\mu^\prime= \cos{\theta} W_\mu - \sin \theta X_\mu, \\
	X_\mu^\prime = \sin{\theta} W_\mu + \cos \theta X_\mu,
\end{array} \right.\eea
where $\theta$ is a small mixing angle and is defined by \bea t_{2\theta}\equiv \tan 2\theta = \fr {-2(w'v+wv')}{v^2 + v'^2 + w^2 + w'^2}.\eea
The mass expressions of the non-hermitian gauge boson can be found in \cite{Van_Loi_2020}. 
In the limit, $v^\prime, w^\prime \ll v \ll w$, two $SU(3)_L$ Higgs triplets
can be written in terms of physical states as follows
\be \rho\simeq \left(
\begin{array}{c}
	G^+_W\\
	\fr{1}{\sqrt{2}}(v+H+i G_Z)\\
	\fr{1}{\sqrt{2}}w^\prime+H'\\
\end{array}
\right),\hs \chi\simeq \left(
\begin{array}{c}
	G^+_X\\
	\fr{1}{\sqrt{2}}v^\prime+G^0_Y\\
	\fr{1}{\sqrt{2}}(w+H_1+iG_{Z'})\\
\end{array}
\right).\label{615d}
\ee 
where $H$ is identified as the $126$ GeV SM like Higgs boson, $H_1, H^\prime$ represent new neutral non SM Higgs bosons, and  $G_{W,X,Y,Z,Z^\prime}$ are the Goldstone bosons associated with the longitudinal components of the $W,X,Y,Z,Z^\prime$ gauge bosons, respectively.
\subsection{Femion mass spectrum}   
The total Yukawa interactions up to six dimensions are given in Ref. \cite{Van_Loi_2020} as follows
\\
\bea
\mathcal{L}_{\text{Yukawa}} =\mathcal{L}^{\text{quark}}_{\text{Yukawa}}+\mathcal{L}^{\text{lepton}}_{\text{Yukawa}}.
\label{Yuka1}\eea
\hs The first term in Eq.(\ref{Yuka1}) contains the quark Yukawa interactions, and it can be written as follows
\bea
\mathcal{L}^{\text{quark}}_{\text{Yukawa}}&=&h^U_{ab}\bar{Q}_{{aL}}\chi^* U_{bR}+h^u_{ab}\bar{Q}_{{aL}}\rho^*u_{bR} + s^u_{ab}\bar{Q}_{{aL}}\chi^*u_{bR} \nn \\
&+&s^U_{ab}\bar{Q}_{{aL}}\rho^*U_{bR}+\frac{h^d_{ab}}{\Lambda} \bar{Q}_{aL} \chi \rho d_{bR}+H.c.
\eea  
After the spontaneous breaking of the $SU(3)_L\times U(1)_N$ electroweak gauge group, the d-quarks gain masses via the following non-renormalizable Yukawa terms:
\bea
[\mathcal{M}_d]_{ab}=\frac{h^d_{ab}}{2\Lambda}(w v - w' v').\eea
The SM $u$-quarks, $u= (u_1,u_2,u_3 )$, and new $U$-quarks, $U=(U_1,U_2,U_3 )$, are mixed via the following mass matrix
\bea
\mathcal{M}_{\text{up}}=\fr{1}{\sqrt{2}}\left( \begin{array}{ccc} 
	h^uv+s^u{v'} & h^U v' +s^U v\\ 
	-h^uw'-s^uw & -h^Uw-s^Uw'  \\
\end{array}\right) = \left(\begin{array}{ccc} 
	M_u & M_{uU}\\ 
	M^T_{uU} & M_U\\
\end{array}\right)\label{eq:massQ}.
\eea
Due to the conditions, $w \gg v \gg w^\prime, v^\prime$, and $h^u,h^U \gg s^u,s^U$, the mass matrix, $\mathcal{M}_{\text{up}}$, allows the implementation 
a type I seesaw mechanism and thus, after a block diagonalization, it is found that the light states, $u^\prime=\left(\begin{array}{ccc} u_1^\prime,u_2^\prime,u_3^\prime \end{array} \right)^T$, and heavy states, $U^\prime=\left(\begin{array}{ccc} U_1^\prime,U_2^\prime,U_3^\prime \end{array} \right)^T$,  are separated as follows:
\bea
u^\prime= u +\left(M_{uU}^* M_u^{-1} \right) U=u+T_u U, \hs U^\prime= U- \left( M_U^{-1}M_{uU}^T\right)u= U-T^\prime_u u
\eea
with $T_u=M_{uU}^* M_U^{-1}$, and $T_u^\prime=M_U^{-1}M_{uU}^T$. The light quarks mix together, and the physical states of light quarks are denoted by
\bea
u^\prime_{L,R}&& = \left(\begin{array}{ccc} u_1^\prime,u_2^\prime,u_3^\prime \end{array} \right)_{L,R}^T= V_{L,R}^u \left(\begin{array}{ccc} u_1,u_2,u_3 \end{array} \right)_{L,R}^T, \nonumber \crn d^\prime_{L,R} && = \left(\begin{array}{ccc} d_1^\prime,d_2^\prime,d_3^\prime \end{array} \right)_{L,R}^T= V_{L,R}^d \left(\begin{array}{ccc} d_1, d_2, d_3 \end{array} \right)_{L,R}^T, \nonumber \crn
U^\prime_{L,R}&& = \left(\begin{array}{ccc} U_1^\prime,U_2^\prime,U_3^\prime \end{array} \right)_{L,R}^T= V_{L,R}^U \left(\begin{array}{ccc} U_1,U_2,U_3\end{array} \right)_{L,R}^T.
\eea

The second term in Eq.(\ref{Yuka1}) contains the Yukawa interactions of leptons,
\bea \mathcal{L}^{\text{lepton}}_{\text{Yukawa}}&=& h^E_{\al b}\bar{\psi}_{\al L} \chi E_{bR} +h^e_{\al b}\bar{\psi}_{\al L}\rho e_{bR}+s^e_{\al b}\bar{\psi}_{\al L}\chi e_{bR}+s^E_{\al b}\bar{\psi}_{\al L} \rho E_{bR} \crn
&&+\fr{h_{1b}^{E}}{\La}\bar{\psi}_{1L}\chi \chi E_{bR}+\fr{h_{1b}^{e}}{\La}\bar{\psi}_{1L}\chi \rho e_{bR} +\fr{s_{1b}^{E}}{\La}\bar{\psi}_{1L}\chi \rho E_{bR}\crn 
&&+\fr{s'^E_{1b}}{\La}\bar{\psi}_{1L}\rho \rho E_{bR}+\fr{s_{1b}^{e}}{\La}\bar{\psi}_{1L}\chi \chi e_{bR}+\fr{{\rm s}_{1b}^{e}}{\La}\bar{\psi}_{1L}\rho \rho e_{bR} \\
&&+ \fr{h_{11}^{\xi}}{\La}\bar{\psi}^c_{1L}\chi \chi \psi_{1L}+\fr{s_{11}}{\La}\bar{\psi}^c_{1L}\chi \rho \psi_{1L}+ \fr{s'_{11}}{\La}\bar{\psi}^c_{1L}\rho \rho \psi_{1L}\crn
&&+\frac{s_{1\al}}{\La^2}\bar{\psi}^c_{1L}\chi \chi\rho \psi_{\al L}+\frac{s'_{1\al}}{\La^2}\bar{\psi}^c_{1L}\chi \rho\rho \psi_{\al L} +H.c. \label{nl2}\eea
From the charged lepton Yukawa interactions, it follows that the SM charged leptons mix with the exotic ones.	 
In the basis, $e_a^\pm, E_a^\pm, \xi^\pm$, the charged lepton mass matrix has the form: 
\bea
\mathcal{M}_l=\left(\begin{array}{ccc} 
	M_{ee} & M_{eE} & M_{e\xi}\\ 
	M^T_{eE} & M_{EE} & M_{E\xi}\\
	M_{e\xi}^T &M_{E\xi}^T & M_{\xi \xi}
\end{array}\right),\label{eq:massCL}
\eea
where,
\bea
[M_{ee}]_{\al b} && \simeq -\frac{1}{\sqrt{2}} h_{\al b}^e v 
+ f^{ee}_{\al b}(v^\prime, w^\prime, s^e_{\al b}(v,w)),\hs
[M_{ee}]_{1 b}  \simeq -\frac{1}{2\sqrt{2}} h_{1 b}^e v + f^{ee}_{1 b}(v^\prime, w^\prime, s^e_{1b}(v,w)), \nonumber
\eea
\bea
[M_{EE}]_{\al b} && \simeq -\frac{1}{\sqrt{2}} h_{\al b}^E w
+ f^{EE}_{\al b}(v^\prime, w^\prime, s^E_{\al b}(v,w)),\hs
[M_{EE}]_{1 b}  \simeq -\frac{1}{2\sqrt{2}} h_{1 b}^E w + f^{EE}_{1 b}(v^\prime, w^\prime, s^{(\prime)E}_{1b}(v,w)), \nonumber
\eea
\bea
[M_{eE}]_{\al b} && \simeq f_{\al b}^{eE}(v^\prime,w^\prime,s^E_{\al b}(v,w)), \hs [M_{eE}]_{1 b}  \simeq f_{1 b}^{eE}(v^\prime,w^\prime,s^{(\prime)E}_{\al b}(v,w)) \nonumber 
\eea
with $\al =2,3$ and
\bea
M_{\xi \xi} =-h_{11}^\xi w+ f^{\xi}(v^\prime), \hs \hs [M_{e\xi}]_{1 b}= f^{e\xi}_{1 b}(w^\prime,v^\prime,s_{1b}^ew,\text{s}_{1b}^ev). 
\eea
The functions, $f^{EE}_{a b},f^{eE}_{a b},f^{e\xi}$, are given in Appendix \ref{app1}. It is very clear to note that the charged lepton mass matrix, $\mathcal{M}_l$, allows the implementation of the type II seesaw mechanism, and the resulting mass eigenstates can be written as follows
\bea
e^\prime && = \left(\begin{array}{ccc} e_1^\prime & e_2^\prime & e_3^\prime \end{array}\right)^T = \left(\begin{array}{ccc} e_1 & e_2  & e_3 \end{array}\right)^T + T_e \left(\begin{array}{cccc} e_1 & e_2 & e_3 & \xi^- \end{array}\right)^T, \nonumber \crn
E^\prime && =\left(\begin{array}{cccc} E_1^\prime & E_2^\prime & E_3^\prime & \xi^{\prime} \end{array}\right)^T = \left(\begin{array}{cccc} E_1 & E_2 & E_3 & \xi \end{array}\right)^T
- T_e^{\prime} \left(\begin{array}{ccc} e_1 & e_2  & e_3 \end{array}\right)^T \eea
with $T_e= \left( \begin{array}{cc} M_{eE}^* &M_{e\xi}^* \end{array}\right) M_{E\xi}^{*-1}$, and $T_e^\prime=M_{E\xi}^{-1}\left( \begin{array}{cc} M_{eE} &M_{e\xi} \end{array}\right)^T$. The light leptons, $e^\prime$, and the heavy states,  $E^\prime$ self-mix, and their physical states are defined via the mixing matrix as follows:
\bea
e^\prime_{L,R} && =\left(\begin{array}{ccc} e_1^\prime & e_2^\prime & e_3^\prime \end{array}\right)_{L,R}^T = V^e_{L,R}\left(\begin{array}{ccc} e_1 & e_2& e_3 \end{array}\right)_{L,R}^T, \crn 
E^\prime_{L,R} && =\left(\begin{array}{cccc} E_1^\prime & E_2^\prime & E_3^\prime & \xi^\prime \end{array}\right)_{L,R}^T = V^E_{L,R}\left(\begin{array}{cccc} E_1 & E_2 & E_3 & \xi\end{array}\right)_{L,R}^T.
\eea
The active neutrinos obtain tiny masses from a combination of type II and III seesaw mechanisms, as follows from the leptonic Yukawa interactions and shown in detail in Ref. \cite{Van_Loi_2020}. We would like to note that the physical neutrino states are related to the flavor states as follows 
\bea
\nu^\prime= \left(\begin{array}{ccc} \nu^\prime_1 & \nu^\prime_2 & \nu^\prime_3 \end{array} \right)_{L,R}^T =V^\nu_{L,R}\left(\begin{array}{ccc} \nu_1 & \nu_2 & \nu_3 \end{array} \right)_{L,R}^T
\eea

\section{New physics effects on charged currents \label{current}}
The interactions of fermions with gauge bosons are derived from the kinetic energy terms of the fermions and have the following form:
\bea
\mathcal{L}_\text{Fermion}^\text{kinetic}=\text{i}\bar{F}\gamma\text{D}_\mu F .
\label{hu}\eea
where $F$ runs on all of the model's fermion multiplets.  We can obtain both neutral and charged currents from Eq. (\ref{hu}). The neutral current has the form
\bea
\mathcal{L}^{\text{N.C}}= -\frac{g}{2{\text{c}}_{W}}
\left\{\bar{f}\gamma^\mu\left[g^Z_V (f) - g^Z_A (f)\gamma_5\right]f Z_\mu 
- \bar{f}\gamma^\mu\left[g^{Z^\prime}_V(f) - g^{Z^\prime}_A(f)\gamma_5\right]f Z^\prime_\mu\right\}, \eea where the vector and axial-vector couplings $g^{Z, Z'}_V (f)$,  $g^{Z, Z'}_A (f)$ can be found in \cite{Van_Loi_2020}.

It is worth noting that the first lepton family transforms as a sextet of $SU(3)_L$, whereas the remaining two families transform as a triplet, freeing the LFUV from both gauge couplings and Yukawa couplings. 
The non-universal interactions of $Z^\prime$, and $X^\pm, Y^{0 (0*)}$ bosons with leptons, are expressed via the following charged currents \cite{Van_Loi_2020}:
\bea 
\mathcal{L}^{\text{C.C}}= \text{J}_W^{-\mu}W^+_\mu+\text{J}_X^{-\mu}X^+_\mu+\text{J}_Y^{0 \mu} Y^0_\mu+\text{H.c}, \label{eff1}
\eea  
where $\text{J}_W^{-\mu}$, $\text{J}_X^{-\mu}$ and $\text{J}_Y^{-\mu}$ are given by: 
\bea
\text{J}_{W}^{-\mu}&&=-\frac{g}{\sqrt{2}}\left\{\bar{\nu}_{{aL}}\ga^\mu e_{{aL}}+\bar{u}_{{aL}}\ga^\mu d_{{aL}} +\sqrt{2}\left(\bar{\xi^+_{{L}}}\ga^\mu \xi^0_{{L}} +\bar{\xi^0_{{L}}}\ga^\mu\xi^-_{{L}}\right)\right\},\\
\text{J}_{X}^{-\mu}&&=-\frac{g}{\sqrt{2}}\left\{\bar{\nu}_{\al {L}}\ga^\mu E_{\al {L}}+\sqrt{2}\left( \bar{\nu}_{1 {L}}\ga^\mu E_{1 {L}}+\bar{\xi^+_{{L}}}\ga^\mu \nu_{1{L}} \right)+\bar{\xi_{{L}}^0}\ga^\mu e_{1{L}}-\bar{U}_{{aL}}\ga^\mu d_{{aL}} \right\}, \label{cc1}\\
\text{J}_{Y}^{0\mu}&&=-\frac{g}{\sqrt{2}}\left\{\bar{e}_{\al {L}} \ga^\mu E_{\al {L}}+\sqrt{2}\left(  \bar{e}_{1{L}} \ga^\mu E_{1{L}} +\bar{\xi^-_{{L}}} \ga^\mu e_{1{L}}\right)+\bar{\xi_{{L}}^0} \ga^\mu \nu_{1{L}}+\bar{U}_{{aL}}\ga^\mu u_{{aL}}\right\} .
\eea

To find the interaction vertexes of fermions-quarks-charged gauge bosons, we have to work with the physical states, such as
\bea
u_{L,R} && = \left(V_{L,R}^u \right)^{-1} u_{L,R}^\prime - T_u \left( V^U_{L,R}\right)^{-1} U_{L,R}^\prime, \crn \nonumber
U_{L,R} && = \left(V_{L,R}^U \right)^{-1} U_{L,R}^\prime + T_u^\prime \left( V^u_{L,R}\right)^{-1} u_{L,R}^\prime, \crn \nonumber
d_{L,R}&=& \left(V_{L,R}^d\right)^{-1} d_{L,R}^\prime, \crn \nonumber
e_{L,R}&& = \left (V_{L,R}^l \right)^{-1}e_{L,R}^\prime -T_e \left( V_{L,R}^E\right)^{-1}E_{L,R}^\prime, \crn \nonumber
E_{L,R}&& = \left (V_{L,R}^E \right)^{-1}E_{L,R}^\prime +T_e^\prime \left( V_{L,R}^l\right)^{-1}e_{L,R}^\prime.
\eea
It is worth noting that $V_{CKM}=V^u_L \left ( V_L^d\right)^\dag $, $U_{PMNS}= V_L^l\left( V_L^\nu \right)^\dag$.
Due to gauge mixing, $W^\pm$ mixed with $X^\pm$, and fermion mixing effects, left-handed SM fermions have anomalous flavor-changing couplings. The part of the Lagrangian describing these interactions, which are obtained  from Eq.(\ref{eff1}), is:
\bea
\delta \mathcal{L}^{C.C} \ni &&-\frac{g}{\sqrt{2}}\left[\left( V_{CKM} \Delta_L^q\right)_{ij}W_\mu^{'+} \bar{u_L^\prime}^i \ga^\mu d^{\prime j}_L + \left(\left(U_{PMNS}\right)^\dag \Delta_L^l\right)_{ij}W_\mu^{'+} \bar{\nu^\prime}^i_L \ga^\mu l^{\prime j}_L\right] \crn 
&& -\frac{g}{\sqrt{2}}\left[ \left(V_{CKM}\Delta_L^{'q} \right)_{ij} X_\mu^{'+}\bar{u^\prime_L}^i \ga^\mu d^{\prime j}_L + \left(\left(U_{PMNS}\right)^\dag \Delta_L^{\prime l}\right)_{ij}X_\mu^{'+} \bar{\nu^\prime}^i_L \ga^\mu l^{\prime j}_L\right]+ H.c,\nn\\
\label{char2}\eea
where, \bea \left(\Delta_L^q\right)_{ij} & =& c_\theta \delta_{ij} +\left( \bar{T_u^\prime}\right)_{ij}s_\theta , \\
 \left( \Delta^{'q}_L \right)_{ij}& =&s_\theta  \delta_{ij} - \left(\bar{T_u^\prime}\right)_{ij}c_\theta,  \\
 \left(\Delta^l_L\right)_{ij} &=& 	\begin{cases}
 	c_\theta  \delta_{ij} - \sqrt{2}s_\theta\left(T_e^\prime\right)_{ij} \hs  \textrm{for } i,j=1 , \\
 	c_\theta  \delta_{ij} - s_\theta\left(T_e^\prime\right)_{ij}\ \hs \hs \textrm{for } i,j=2,3 , 
 \end{cases}  \\ 
\left(	\Delta^{'l}_L\right)_{ij} &=& 
\begin{cases}
	s_\theta  \delta_{ij} + \sqrt{2}c_\theta\left(T_e^\prime\right)_{ij} & \textrm{for } i,j=1 , \\
	s_\theta  \delta_{ij} + c_\theta\left(T_e^\prime\right)_{ij} & \textrm{for } i,j=2,3 .
\end{cases} \eea 
\section{The flavor non-universality effective Hamiltonian in $u^i - d^j$ transitions  \label{Eff}}
The contributions of charged-currents (\ref{char2}) to lepton-flavor non-universal processes, such as the $u^i-d^j$ transition, are contained in the effective Hamiltonian  
\bea
\mathcal{H}_{eff}= \left[ \text{C}_{\nu_a e_b}^{u_id_j}\right]\left( \bar{u}^\prime_{iL}\ga^\mu d_{jL}^{'}\bar{\nu}_{aL}^\prime \ga_\mu e^\prime_{bL} \right).
\eea
At tree level, the Wilson coefficients (WCs), $\left[ \text{C}_{\nu_a e_b}^{u_id_j}\right]_{\text{tree}}$, are separated as follows
\bea  \left[ \text{C}_{\nu_a e_b}^{u_id_j}\right]_{\text{tree}}= \left[ \text{C}_{\nu_ae_c}^{u_id_k}\right]_{\text{SM}}\left(\delta\left[ \text{C}^{u_k d_j}_{\nu_ce_b}\right]_{W_\mu^{'}}+\delta\left[ \text{C}^{u_k d_j}_{\nu_c e_b}\right]_{X_\mu^{'}}\right)\eea with
\bea
\left[ \text{C}_{\nu_ae_c}^{u_id_k}\right]_{\text{SM}}&& =\frac{4G_F}{\sqrt{2}}\left(U_{PMNS}\right)^\dag_{ac}\left( V_{CKM}\right)_{ik}, \crn \\
\delta\left[ \text{C}_{\nu_c e_b}^{u_kd_j}\right]_{W_\mu^{'}} && = \left(\Delta_L^q\right)_{kj}\left(\Delta_L^l\right)_{cb},\crn \\
\delta\left[ \text{C}_{\nu_c e_b}^{u_kd_j}\right]_{X_\mu^{'}} && = \frac{m_W^2}{m_X^2}\left(\Delta_L^{\prime q}\right)_{kj}\left(\Delta_L^{\prime l}\right)_{cb}.
\eea

\begin{figure}[H]
	{\includegraphics[width=12cm]{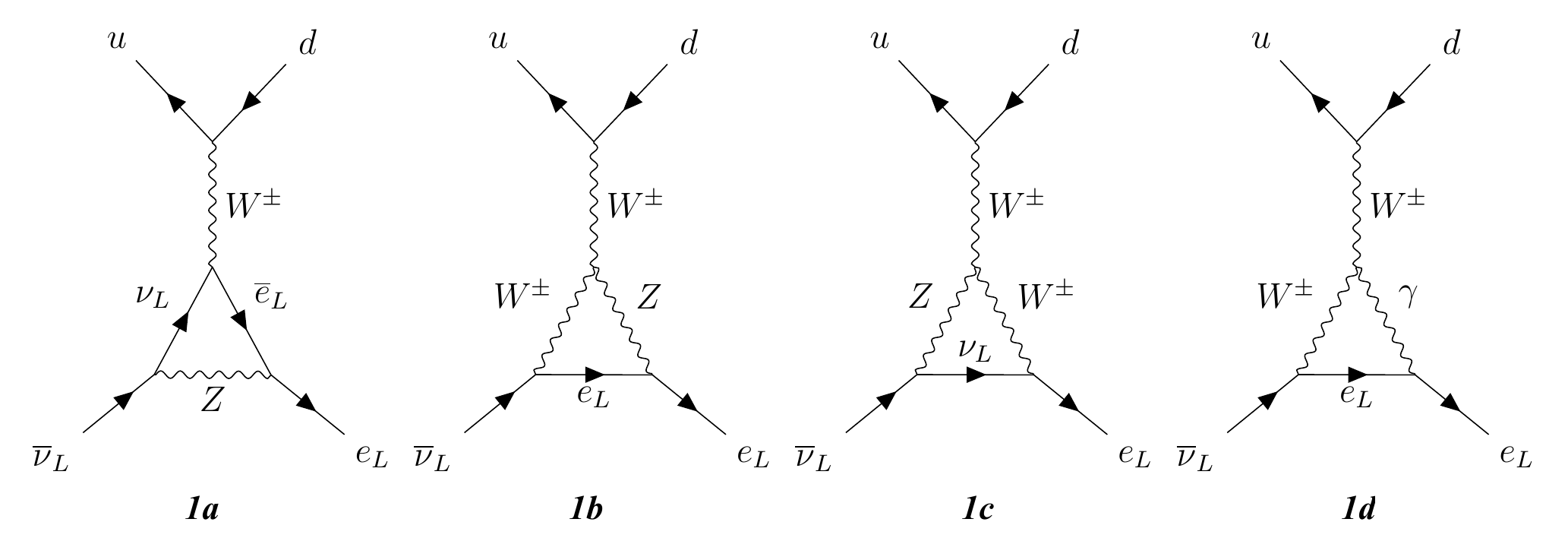}}
	{\caption{Penguin diagrams that are obtained from the SM interactions}
		\label{penguinSM}}
\end{figure}

\begin{figure}[H]
	{\includegraphics[width=12cm]{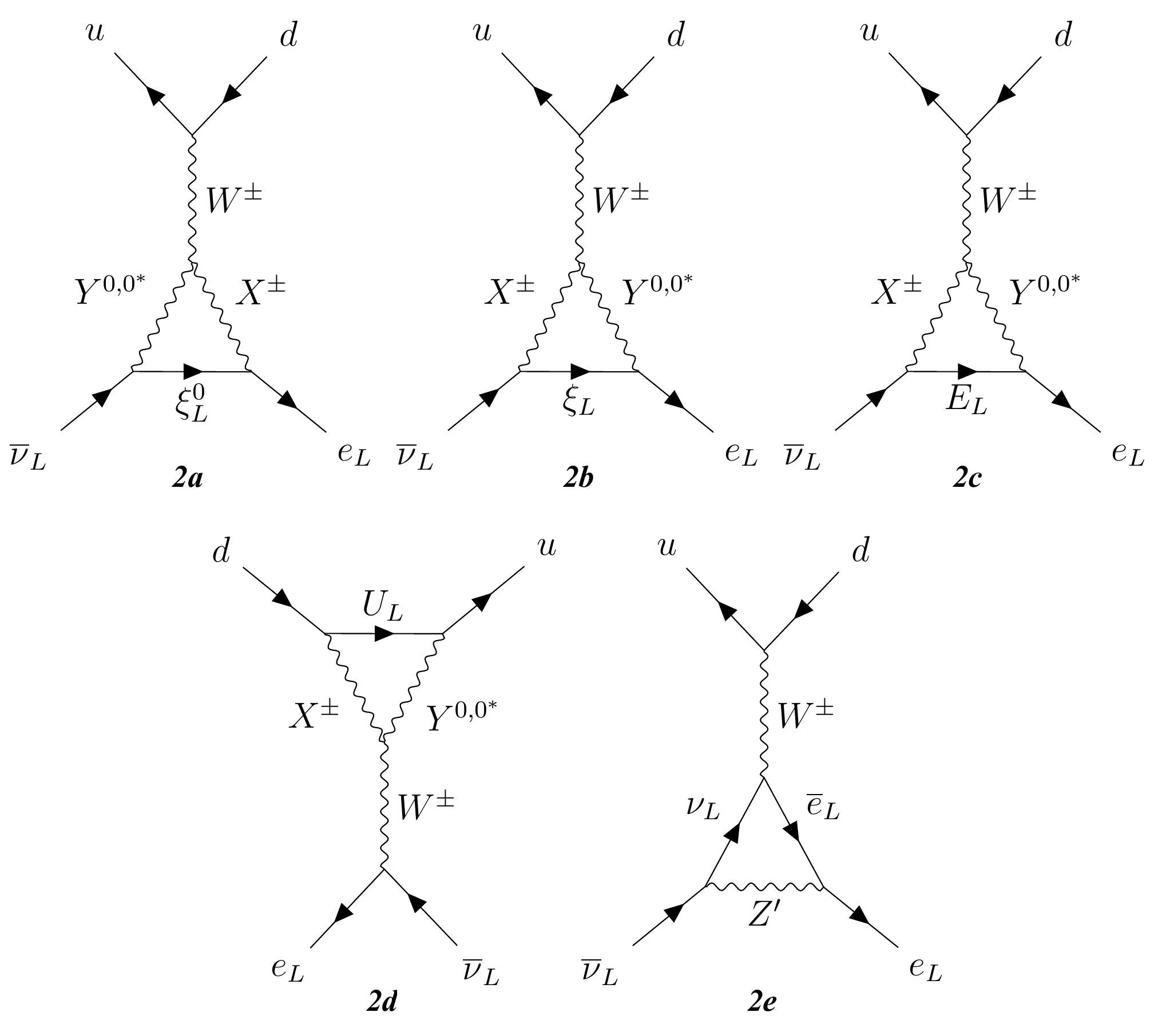}}
	{\caption{Penguin diagrams that are obtained from the new interactions}
		\label{penguinNP}}
\end{figure}

\begin{figure}[H]
	{\includegraphics[width=12cm]{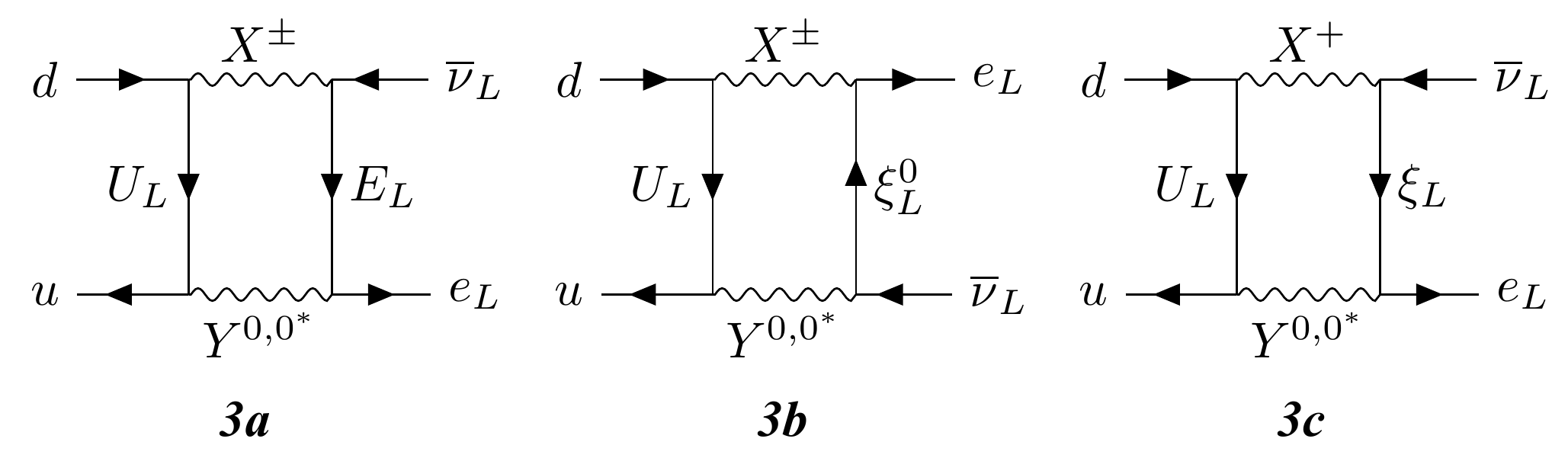}}
	{\caption{Box diagram}
		\label{box}}
\end{figure}

The intensity of the new interactions is of the order of $\frac{v^{\prime }, w^{\prime }}{v,w} \simeq (\varepsilon^2)$, which implies that NP contributions arising from the tree-level exchange of heavy vector bosons are very suppressed. The non-universal interactions of the SM leptons and new leptons with new gauge bosons also generate the four-fermion interactions via the one loop level box and penguin diagrams, which are shown in Figs. (\ref{penguinSM}), (\ref{penguinNP}), (\ref{box}).

Including one loop correction, the WCs, $\text{C}^{u_id_j}_{\nu_a e_b}$, can be separated as follows
\bea
\text{C}^{u_id_j}_{\nu_a e_b}=\left[\text{C}^{u_id_j}_{\nu_a e_b}\right]_{\text{tree}} +\left[\text{C}^{u_id_j}_{\nu_a e_b}\right]_{\text{penguin}}+\left[\text{C}^{u_id_j}_{\nu_a e_b}\right]_{\text{box}}.
\eea 
We split the penguin diagrams' contribution into two components 
\bea
\left[\text{C}^{u_id_j}_{\nu_a e_b}\right]_\text{penguin}=\left[\text{C}^{u_id_j}_{\nu_a e_b}\right]_\text{penguin}^{\text{SM}}+\left[\text{C}^{u_id_j}_{\nu_a e_b}\right]_\text{penguin}^{\text{NP}}.
\eea
The first SM's contribution is denoted by the $\left[\text{C}^{u_id_j}_{\nu_a e_b}\right]_\text{penguin}^{\text{SM}}$ coefficient , which is written as
\bea
\begin{multlined}
	\left[\text{C}^{u_id_j}_{\nu_a e_b}\right]_\text{penguin}^{\text{SM}}= \frac{4G_F}{\sqrt{2}}\frac{ 3g^2}{512 \pi^2}\times \left\{\frac{1}{m^2_Z-m^2_W}
	\left[\text{C}^{u_id_j}_{\nu_a e_b}\right]^{WWZ} \right.\\
	\left.+\frac{1}{m^2_{e_b}-m^2_{\nu_a}}\left[\text{C}^{u_id_j}_{\nu_a e_b}\right]^{We\nu}_Z +\frac{1}{m_W^2}\left[\text{C}^{u_id_j}_{\nu_a e_b}\right]^{WW\ga}
	\right\},
\end{multlined}
\eea
where the coefficient $\left[\text{C}^{u_id_j}_{\nu_a e_b}\right]^{WWZ}$ is the contribution of the penguin diagrams with label (1b,1c) illustrated in  Figs. (\ref{penguinSM}) and takes the form:
\bea
\left[\text{C}^{u_id_j}_{\nu_a e_b}\right]^{WWZ} && =  \left[\text{C}^{u_id_j}_{\nu_a e_b}\right]_\text{1b}^{WWZ}+
\left[\text{C}^{u_id_j}_{\nu_a e_b}\right]_\text{1c}^{WWZ}
\nonumber \\ && =4\left(U_{PMNS} \right)^\dag_{ab}\left\{\left(2s_W^2-1 \right)\Ga^{WZe_b}+ \Ga^{WZ\nu_a} \right\}\left(V_{CKM} \right)_{ij}.
\label{SMpenguin1}\eea 
According to the penguin diagram with label (1a) shown in Fig.(\ref{penguinSM}), the coefficient $\left[\text{C}^{u_id_j}_{\nu_a e_b}\right]_Z^{We\nu}$ is calculated by
\bea
\left[\text{C}^{u_id_j}_{\nu_a e_b}\right]_Z^{We\nu}= \left(t_W^2- 1\right)\left( U_{PMNS}\right)^\dag_{ab} \Ga^{W\nu_a e_b}_Z\left(V_{CKM} \right)_{ij}. \label{SMpenguin2}
\eea
The penguin diagram with label (1d) is the last contribution made by the SM. It is given as 
\bea
\left[\text{C}^{u_id_j}_{\nu_a e_b}\right]^{WW\ga}= 8s_W^2 \left(U_{PMNS} \right)^\dag_{ab} \Ga^{W\ga e_b}_Z \left(V_{CKM} \right)_{ij}.
\label{SMpenguin3}\eea
When the mixing effect of new fermions and the SM fermions is ignored, the penguin diagram gives new contributions,
\bea
\left[\text{C}^{u_id_j}_{\nu_a e_b}\right]_\text{penguin}^{\text{NP}}=\left[\text{C}^{u_id_j}_{\nu_a e_b}\right]_{Z^\prime}^{We\nu}+\left[\text{C}^{u_id_j}_{\nu_a e_b}\right]^{WXY}_{\text{penguin}}
\eea
with
\bea
\left[\text{C}^{u_id_j}_{\nu_a e_b}\right]_{Z^\prime}^{We\nu}&& = \frac{4G_F}{\sqrt{2}}\frac{ 3g^2}{512 \pi^2}  \frac{1}{m_{e_b}^2-m_{\nu_a}^2}\frac{1}{c_W^2 \left( 1+2c_{2W}\right)} \crn \nonumber && \times\left[c_{2W}^2 \left(V_L^\nu \right)_{1a}\left( V_L^l\right)^\dagger_{1b} +\left(V_L^\nu \right)_{\al a}\left( V_L^l\right)^\dagger_{\al b}\right] \Ga^{W\nu_a e_b}_{Z^\prime} \left(V_{CKM}\right)_{ij},
\crn \nonumber \left[\text{C}^{u_id_j}_{\nu_a e_b}\right]^{WXY}_{\text{penguin}} && = \frac{4G_F}{\sqrt{2}}\frac{ 3g^2}{128 \pi^2}  \frac{1}{m_{X}^2-m_{Y}^2} \left\{\left[\text{C}^{u_id_j}_{\nu_a e_b}\right]^{WXY}_{2a}+\left[\text{C}^{u_id_j}_{\nu_a e_b}\right]^{WXY}_{2b}+\left[\text{C}^{u_id_j}_{\nu_a e_b}\right]^{WXY}_{2c}+\left[\text{C}^{u_id_j}_{\nu_a e_b}\right]^{WXY}_{2d} \right\},
\eea
where  
\bea
\left[\text{C}^{u_id_j}_{\nu_a e_b}\right]_\text{2a}^{WXY}&& = \left(V^\nu_L\right)_{1a}\left(V^l_L\right)^\dag_{1b} \Ga^{XY \xi^0} \left(V_{CKM} \right)_{ij},\\ \nonumber \left[\text{C}^{u_id_j}_{\nu_a e_b}\right]_\text{2b}^{WXY}&& = 2\left(V^\nu_L\right)_{1a}\left(V^l_L\right)^\dag_{1b} \Ga^{XY \xi} \left(V_{CKM} \right)_{ij}, \\ \nonumber \left[\text{C}^{u_id_j}_{\nu_a e_b}\right]_\text{2c}^{WXY}&& =\sum_{c=1}^3 \mathcal{G}^{\nu_a E_c X} \Ga^{XYE_c} \left( \mathcal{G}^{l_b E_c Y}\right)^\dag\left(V_{CKM} \right)_{ij}, \\ \nonumber
\left[\text{C}^{u_id_j}_{\nu_a e_b}\right]_\text{2d}^{WXY}&&= \left(U_{PMNS} \right)_{ab} \sum_{c=1}^3 \left(V_L^u \left(V_L^{U} \right)^ \dag\right)_{ic} \Ga^{XY U_c} \left(V_L^{U}\left(V_L^{d}\right)^\dag\right)_{cj}.
\eea 
The couplings, $\mathcal{G}^{\nu_a E_c X}, \mathcal{G}^{e_b E_c Y} $ are determined  as follows:
\bea
\mathcal{G}^{\nu_a E_c X} && = \left(V_L^\nu\right)_{a\al} \left(V_L^E\right)^\dag_{\al c} +\sqrt{2} \left(V_L^\nu\right)_{a1} \left(V_L^E\right)^\dag_{1c}, \\ \nonumber
\left( \mathcal{G}^{e_b E_c Y}\right)^\dag && = \left(V_L^E \right)_{c \al} \left(V_L^l\right)_{\al b}^\dag+ \sqrt{2} \left( V_L^E \right)_{c1}\left(V_L^l\right)_{1b}^\dag,
\eea
where $\al =2,3,$ and $\Ga^{ABC}$ is given in Appendix. (\ref{app2}).

The box diagrams are shown in the Fig.(\ref{box}) and their contribution to the WCs are as follows
\bea
\left[\text{C}^{u_id_j}_{\nu_a e_b}\right]_\text{box}=- \frac{4 G_F}{\sqrt{2}}\frac{51g^2}{64 \pi^2}\frac{m_W^2}{m_X^2-m_Y^2}\left \{\left[\text{C}^{u_id_j}_{\nu_a e_b}\right]_\text{box}^{E}+\left[\text{C}^{u_id_j}_{\nu_a e_b}\right]_\text{box}^{\xi^0}+\left[\text{C}^{u_id_j}_{\nu_a e_b}\right]_\text{box}^\xi \right\},
\eea
where
\bea
\left[\text{C}^{u_id_j}_{\nu_a e_b}\right]_\text{box}^{E} &=& \sum_{l=1}^3 \sum_{c=1}^3 \left(V^u_L \left(V^U_L\right)^\dag \right)_{il}\left(V^U_L \left(V^d_L\right)^\dag \right)_{lj} \Ga^{U_lE_c}\mathcal{G}^{\nu_a E_c X} \left( \mathcal{G}^{l_b E_c Y}\right)^\dag, 
\\ \nonumber
\left[\text{C}^{u_id_j}_{\nu_a e_b}\right]_\text{box}^{\xi^0} &=& \sum_{l=1}^3 \left(V^u_L \left(V^U_L\right)^\dag \right)_{il}\left(V^U_L \left(V^d_L\right)^\dag \right)_{lj} \Ga^{U_l \xi^0}\left(V_L^\nu \right)_{a1} \left( V_L^l\right)_{1b}^\dag, \\ \nonumber 
\left[\text{C}^{u_id_j}_{\nu_a e_b}\right]_\text{box}^{\xi} &=&2 \sum_{l=1}^3 \left(V^u_L \left(V^U_L\right)^\dag \right)_{il}\left(V^U_L \left(V^d_L\right)^\dag \right)_{lj} \Ga^{U_l \xi}\left(V_L^\nu \right)_{a1} \left( V_L^l\right)_{1b}^\dag.
\eea

The functions $\Ga^{U_l E_c}$, $\Ga^{U_l\xi^0}$, $\Ga^{U_l \xi}$ are defined respectively in Appendix \ref{app3}

\section{Studying several observables connected to the flavour non-universality of interactions \label{Phys}}
\subsection{ $b \to c$ transitions}
Let's look at the NP effects in $b $ to $c $ transitions. Both the exclusive and inclusive ratios, 
$\text{R}(\text{D}^{(*)})$, $\text{R}(\text{X}_\text{c})$, are taken into account. The ratios, $\text{R}(\text{D}^{(*)}),\text{R}(\text{X}_\text{c})$, which come from NP in the form of the Wilson coefficients, are presented in  \cite{Boucenna:2016qad} as: 
\bea
\text{R}(\text{D}^{(*)}) & \equiv&  \frac{\Gamma \left(\text{B} \to \text{D}^{(*)}\tau \bar{\nu} \right)}{\Ga \left(\text{B} \to \text{D}^{(*)}\l \bar{\nu} \right)} =  \frac{ \sum_k|\text{C}^{cb}_{3j}|^2}{\sum_k \left(|\text{C}^{cb}_{1k}|^2+|\text{C}^{cb}_{2k}|^2\right)} \times\left[\frac{\sum_k \left(|\text{C}^{cb}_{1k}|^2+|\text{C}^{cb}_{2k}|^2\right)}{ \sum_k|\text{C}^{cb}_{3k}|^2}\right]_ {\text{SM}}\times \text{R}(\text{D}^{(*)})_{\text{SM}}, \nonumber \crn
\text{R}(\text{X}_\text{c}) &\equiv &  \frac{\Gamma \left(\text{B} \to \text{X}_{\text{c}}\tau \bar{\nu} \right)}{\Ga \left(\text{B} \to \text{X}_{\text{c}}l \bar{\nu} \right)}  = \frac{\sum_k|\text{C}^{cb}_{3k}|^2}{\sum_k |\text{C}^{cb}_{1k}|^2}  \times\left[\frac{\sum_k| \text{C}^{cb}_{1k}|^2}{ \sum_k|\text{C}^{cb}_{3k}|^2}\right]_ {\text{SM}}\times \text{R}(\text{X}_{\text{c}})_{\text{SM}}.
\eea
with $k = 1,2,3$ is the generation index of the leptons.\\
Eq.(\ref{RDSM}) determines the ratio $\text{R}(\text{D}^{(*)})_{\text{SM}}$, and  the ratio $\text{R}(\text{X}_{\text{c}})_{\text{SM}}=0.223(5)$, which is reported in \cite{PhysRevD.90.034021}. 
Furthermore, the experimental value for the inclusive ratio, $\text{R}(\text{X}_\text{c})$, is determined as follows
\bea
\text{R}(\text{X}_{\text{c}})_{\text{exp}}= 0.222(22),
\eea
and the average values of the measurements, $\text{R}_{\text{D}},\text{R}_{\text{D}^*} $, are given in Eq.(\ref{tnph1}).
The discrepancy between the measured values of $\text{R}_{\text{D}},\text{R}_{\text{D}^*} $ and their respective SM predictions is an indication of the presence of NP, whose effects are encoded in the NP Wilson coefficients. In contrast, the experimental result is in slight tension with the SM prediction of $\text{R}(\text{X}_{\text{c}})$. NP effects in the $\text{R}(\text{X}_{\text{c}})$ lead to new stringent constraints on the NP parameters.
In the next study, we fit the parameter space of the considered model by using the data on the observables, $\text{R}_{\text{D}},\text{R}_{\text{D}^*}$.
As was already indicated, the NP Wilson coefficients depend not only on the SM parameters but also on the new parameters such as the mixing matrices, $V_L^d, V_L^u, V_L^U, V_L^E, V_L^l, V_L^\nu$, the new particle masses, $m_{X},m_{Y}, m_{Z^\prime}, m_{E_i}, m_{\xi}, m_{\xi^0}$, and $m_{U_i}$. To perform a numerical study, we use the SM parameters reported in \cite{Workman:2022ynf} and the new parameters are assumed as follows: 
\begin{itemize}
	\item The lepton and quark mixing matrices take the following form:
	\bea
	V_L^l= V_L^u= V_L^U= V_L^E=\text{Diag}\left(1,1,1 \right), \hs V_L^\nu= U_{\text{PMNS}}, \hs V_L^d= V_{\text{CKM}}. \hs 
	\eea 
	this corresponds to the choice of basis where the up type quark and charged lepton mass matrices are diagonal so that the observed quark and lepton mixings only arise from the down type quark and neutrino sectors, respectively. 
	\item To satisfy the LHC constraints \cite{Workman:2022ynf}, the new gauge boson masses were selected as follows: $m_{Z'}=4500 \text{GeV}, m_{X} = 4100 \text{GeV}, m_Y^2 =m_X^2+m_W^2.$
	\item Without loss of generality, we investigate the mass hierarchy of new fermions according to four scenarios:
	\begin{itemize}
		\item The mass of three new leptons, $E_i$, is $m_{E_1}=m_{E_2}=m_{E_3}$, and the mass of three exotic quarks is, $m_{U_1}=m_{U_2}=m_{U_3}$.
		\item Both kinds of new quarks, leptons, have the following normal mass hierarchy, denoted by the symbol $(\text{E}_\text{n} \text{U}_\text{n})$:  $\frac{m_{E_1}}{m_{E_2}}=\frac{m_e}{m_\mu},\frac{m_{E_1}}{m_{E_3}}=\frac{m_e}{m_\tau}$,  $\frac{m_{U_1}}{m_{U_2}}=\frac{m_u}{m_c}, \frac{m_{U_1}}{m_{U_3}}=\frac{m_u}{m_t}$.
		\item Both kinds of new quarks, leptons, have the following inverted mass hierarchy, denoted by the symbol $(\text{E}_\text{i} \text{U}_\text{i})$: $\frac{m_{E_1}}{m_{E_2}}=\frac{m_\mu}{m_e}, \frac{m_{E_1}}{m_{E_3}}=\frac{m_\tau}{m_e}$,  $\frac{m_{U_1}}{m_{U_2}}=\frac{m_c}{m_u}, \frac{m_{U_1}}{m_{U_3}}=\frac{m_t}{m_u}$.
		\item The new leptons have a normal mass hierarchy, but the exotic quarks have an inverted mass hierarchy, denoted by the symbol $(\text{E}_\text{n} \text{U}_\text{i})$: $\frac{m_{E_1}}{m_{E_2}}=\frac{m_e}{m_\mu},\frac{m_{E_1}}{m_{E_3}}=\frac{m_e}{m_\tau}$, $\frac{m_{U_1}}{m_{U_2}}=\frac{m_c}{m_u},\frac{m_{U_1}}{m_{U_3}}=\frac{m_t}{m_u}$, and vice versa $(\text{E}_\text{i} \text{U}_\text{n})$.
	\end{itemize}
\end{itemize}

\begin{figure}[h]
	\includegraphics[width=15cm,height=6cm]{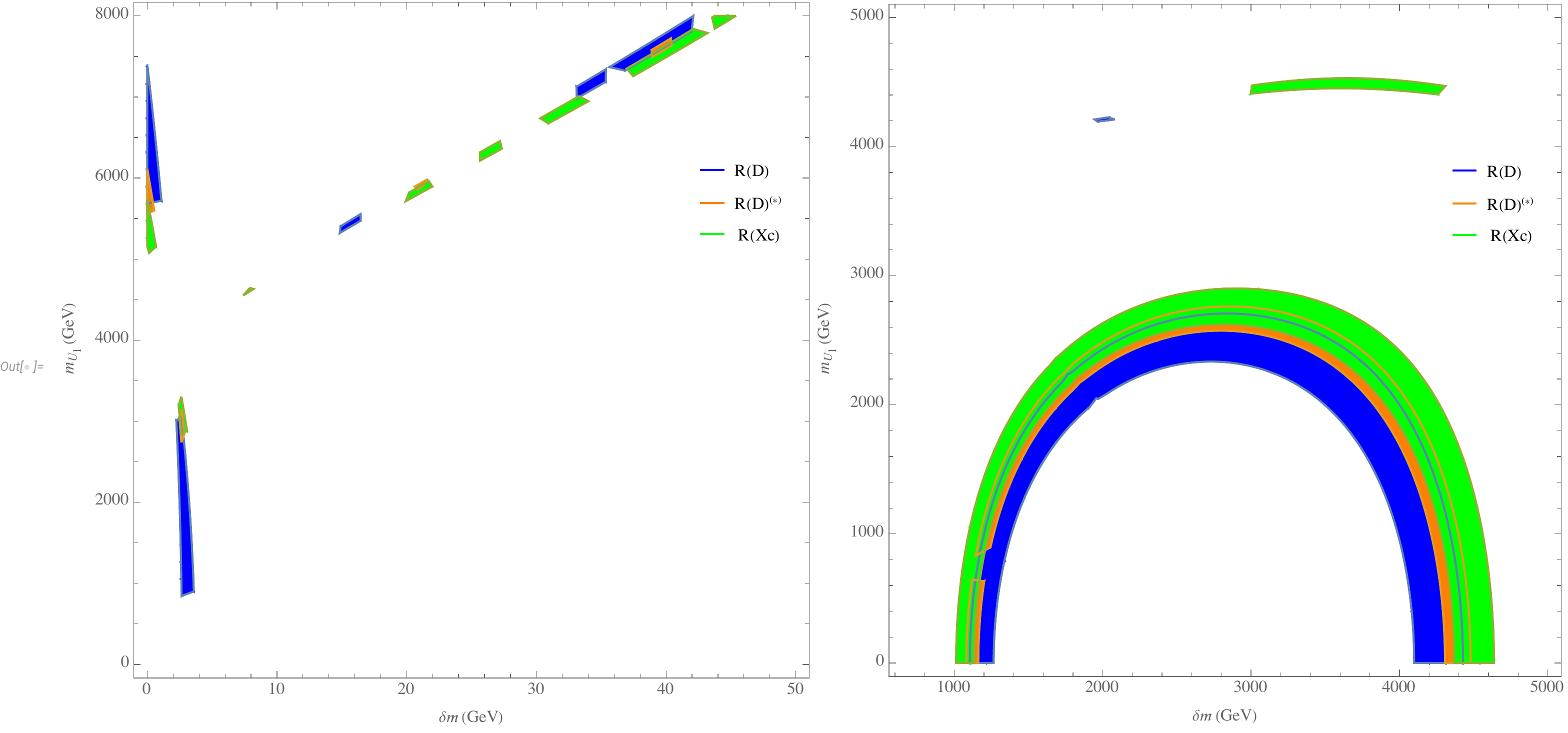}
	\caption{The allowed parameter space in the $\delta m-m_{U_1}$ plane consistent with the experimental constraints of the $\text{R}(\text{D}), \text{R}(\text{D}^{*}), \text{R}(\text{X}_c)$ observables for the first scenario. Here we have set $m_{E_1}=m_{E_2}=m_{E_3}$ and $m_{U_1}=m_{U_2}=m_{U_3}$, $m_\xi =m_{E_1}+\delta m, m_{\xi^0}=m_{E_1}-\delta m$. The blue, orange, and green regions are consistent with the experimentally allowed ranges of the $\text{R}(\text{D}), \text{R}(\text{D}^{*})$ and $\text{R}(\text{X}_c)$ observables, respectively.} 
\label{RDDX1}
\end{figure}
In the first scenario, shown in the plots of Fig. (\ref{RDDX1}), we display the allowed blue, orange, and green regions of parameter space in the $\delta m-m_{U_1}$ plane consistent with the experimental constraints of the $\text{R}(\text{D}), \text{R}(\text{D}^{*}), \text{R}(\text{X}_c)$ observables, respectively. We assumed: $m_{E_1}=m_{E_2}=m_{E_3}$ and $m_{U_1}=m_{U_2}=m_{U_3}$, $m_\xi =m_{E_1}+\delta m, m_{\xi^0}=m_{E_1}-\delta m$. 
As follows from the plots of Fig. (\ref{RDDX1}), we found that the experimental values of the $\text{R}(\text{D}), \text{R}(\text{D}^{*}), \text{R}(\text{X}_c)$ observables can be successfully accommodated in two regions of $\delta m$, one of the order of few GeV and the other one of the order of few TeV. As follows from the plot in the right panel of Fig. (\ref{RDDX1}), we found an upper limit for the exotic quark mass $m_{U_1}$ less than $5$ TeV.
\begin{figure}[H]
	\includegraphics[width=15cm,height=5cm]{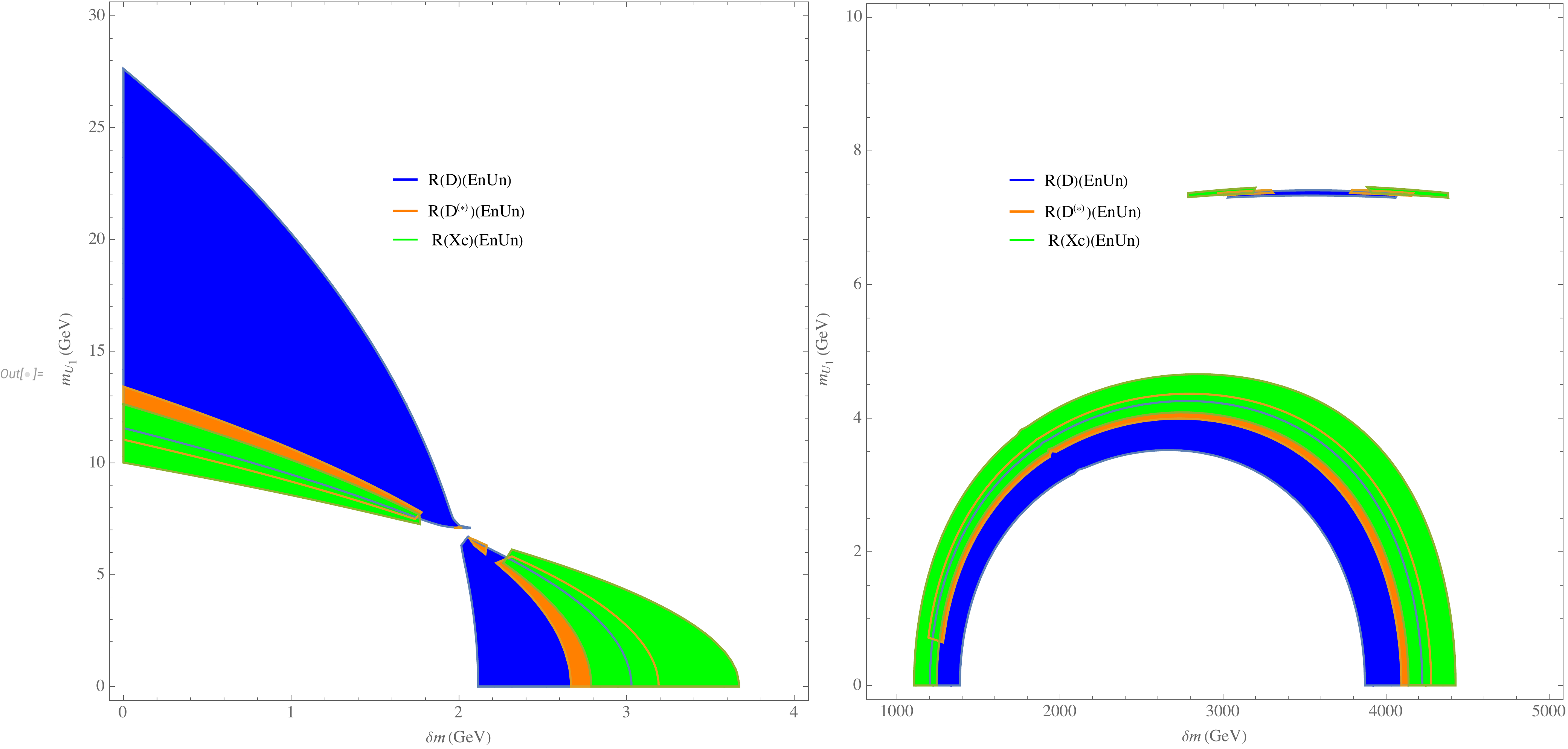}
	\includegraphics[width=15cm,height=5cm]{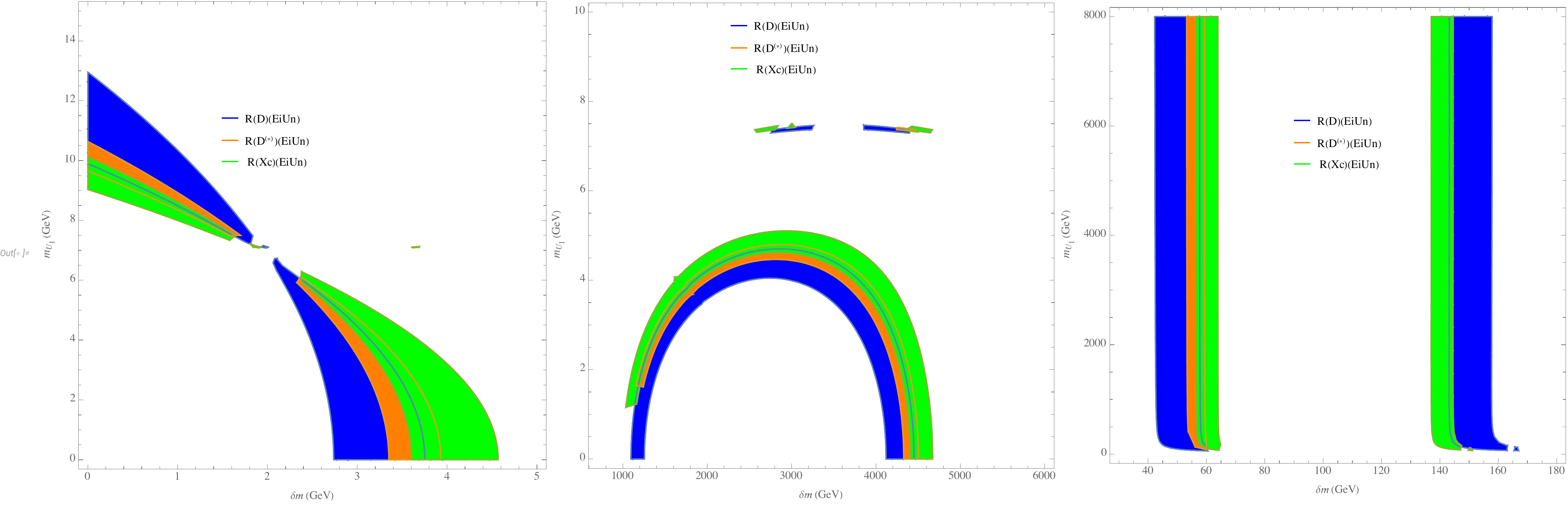}
	\includegraphics[width=15cm,height=5cm]{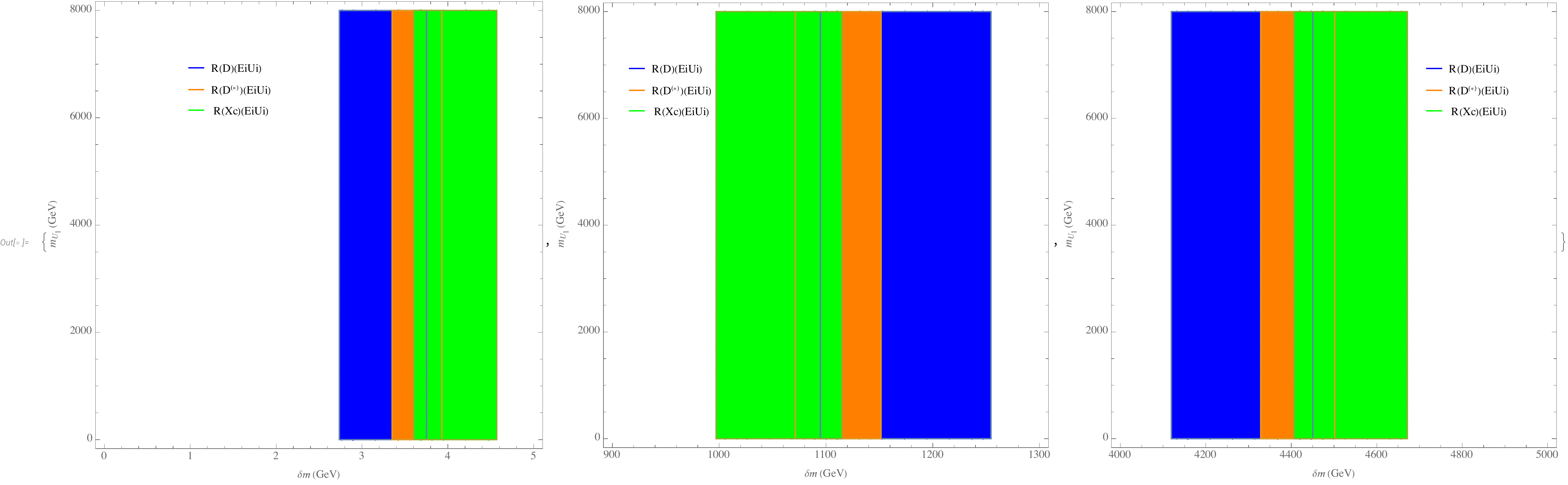}
	\includegraphics[width=15cm,height=5cm]{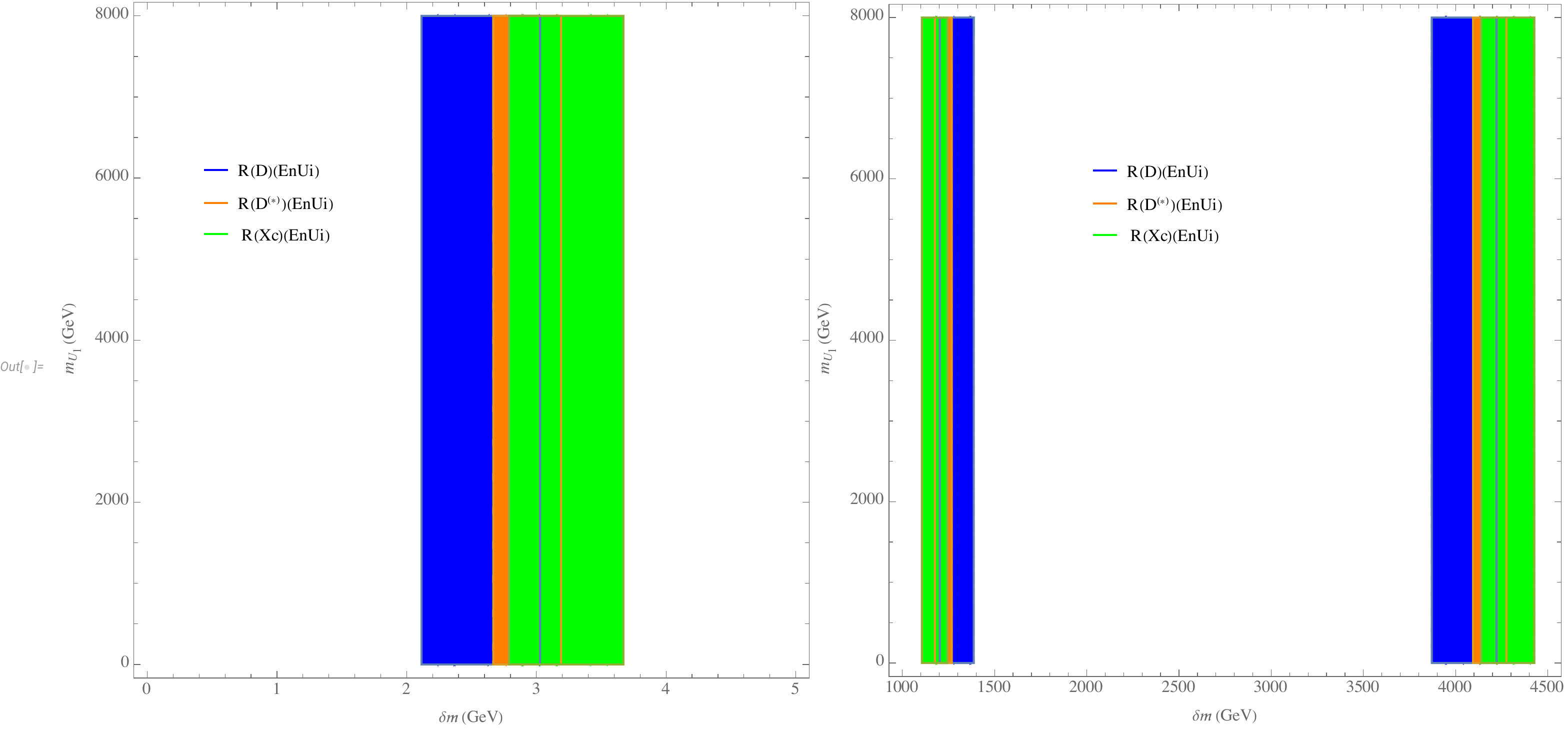}
	\caption{The allowed parameter space in the $\delta m-m_{U_1}$ plane is consistent with the experimental constraints of the ratios, $\text{R}(\text{D}), \text{R}(\text{D}^{*})$, and $\text{R}(\text{X}_c)$ in the scenarios of $\text{E}_\text{n} \text{U}_\text{n}, \text{E}_\text{i} \text{U}_\text{i}$, $\text{E}_\text{i} \text{U}_\text{n}, \text{E}_\text{n} \text{U}_\text{i}$.}  
	\label{RDDX2}
\end{figure}
In the scenarios, both quarks and leptons have the same mass hierarchy, $\text{E}_\text{n} \text{U}_\text{n}$ or $\text{E}_\text{i} \text{U}_\text{i}$ or either quark or lepton has the inverted mass hierarchy, $\text{E}_\text{i} \text{U}_\text{n}, \text{E}_\text{n} \text{U}_\text{i}$, can  find a parameter space of  $m_{U_1}$, $ \delta m$, which can successfully accommodate the experimental values of the $\text{R}(\text{D}), \text{R}(\text{D}^{*}), \text{R}(\text{X}_c)$ observables. We display the allowed region of parameter space in the $\delta m-m_{U_1}$ plane consistent with the experimental constraints in Fig.(\ref{RDDX2}). 
The $\delta m$ is constrained by the experimental values of these ratios, as shown in Fig.(\ref{RDDX2}), and the mass hierarchy of new leptons and quarks. All scenarios, $\text{E}_\text{i} \text{U}_\text{i},\text{E}_\text{n} \text{U}_\text{n}$, $\text{E}_\text{i} \text{U}_\text{n}, \text{E}_\text{n} \text{U}_\text{i}$, accommodate the $\delta m$ to achieve the electroweak enegy scale and the TeV energy scale. The parameter space distribution shape depends on the mass hierarchy of the exotic quarks . For the case of $ \text{E}_\text{i} \text{U}_\text{n}, \text{E}_\text{n} \text{U}_\text{n}$, the allowable region of $\delta m$ at the electroweak scale can range from several GeV to several tens of GeV, the allowed  regions at the TeV scale are restricted by a curved surface, creating a limit on the exotic quark mass. For the case, $\text{E}_\text{i} \text{U}_\text{i}, \text{E}_\text{n} \text{U}_\text{i}$, the parameter space of $\delta m$ only reaches up to a few GeV, and the allowed energy scale at the TeV is part of the plane bounded by lines that the $\delta m$ is constant in the $\delta m- m_{U_1}$. This is equivalent to not creating the limit on the exotic quark.
 
\subsection{$s \to u$  transitions}
\begin{figure}[H]
	\includegraphics[width=15cm,height=5cm]{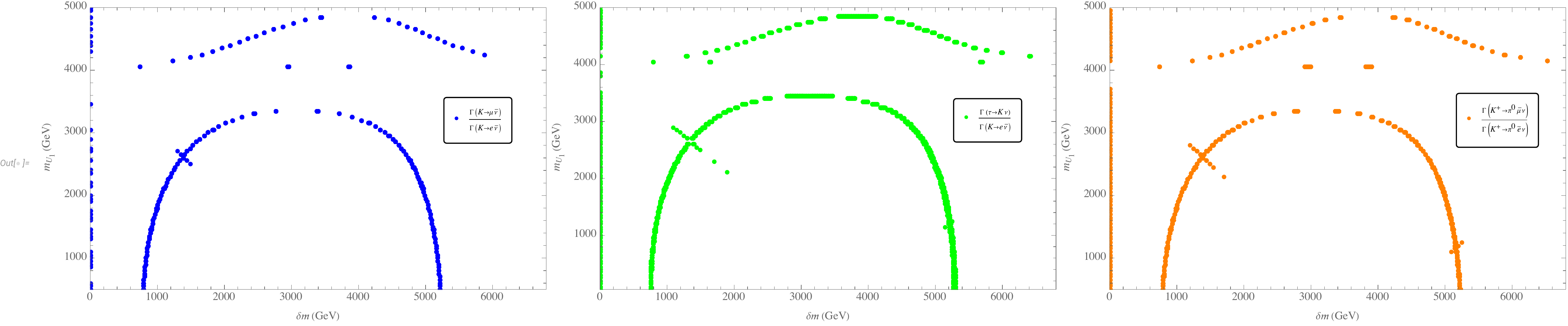}
	\includegraphics[width=15cm,height=5cm]{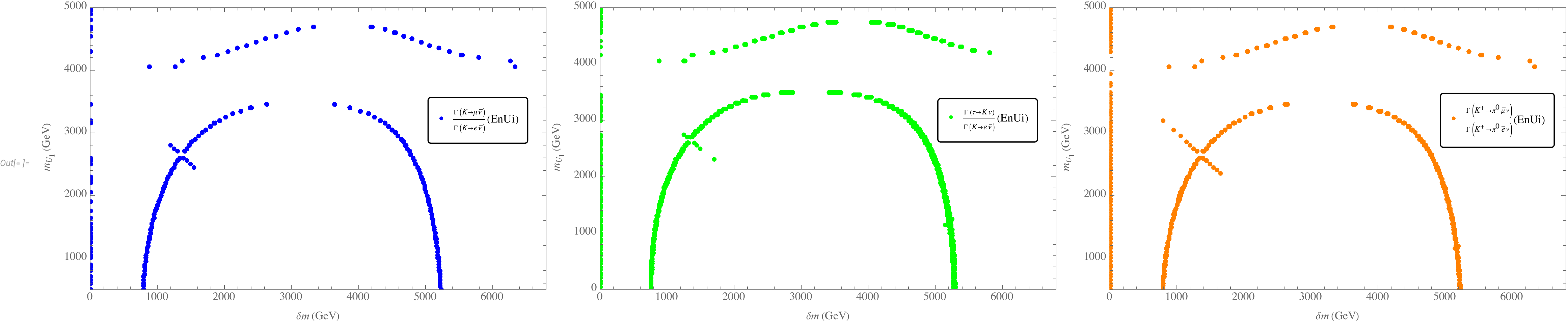}
	\includegraphics[width=15cm,height=5cm]{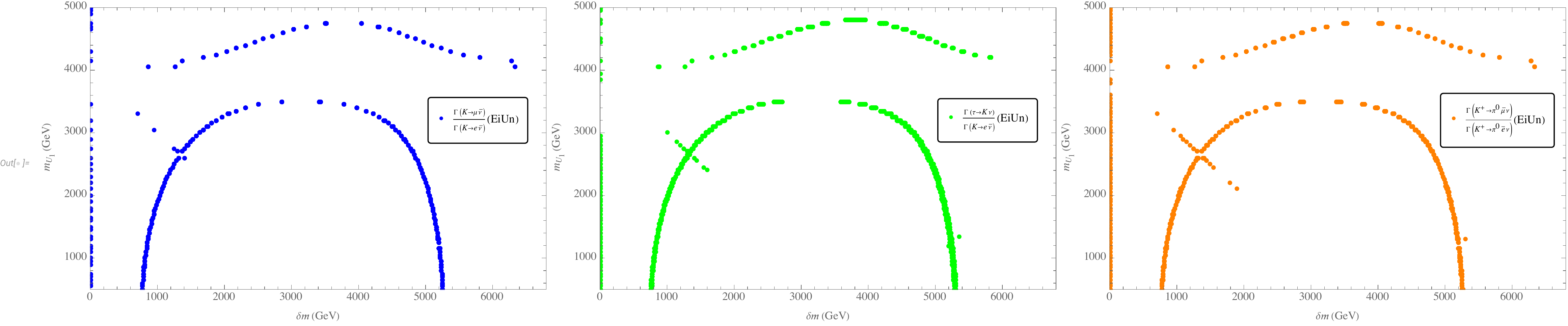}
	\caption{The ratios,$\frac{\Gamma (\text{K} \to \mu \bar{\nu})}{\Gamma(\text{K}\to e \bar{\nu})}, \frac{\Gamma (\tau \to \text{K} \nu)}{\Gamma(\text{K}\to e \bar{\nu})}$ and $\frac{\Gamma (\text{K}^+ \to \pi^0 \bar{\mu}\nu)}{\Gamma(\text{K}^+\to \pi^0 \bar{e} \nu)}$ contoured as a function of $\delta m, m_{U_1}$. The panels from top to bottom correspond to the cases: $m_{E_1}=m_{E_2}=m_{E_3}$, $m_{U_1}=m_{U_2}=m_{U_3}$; $\text{E}_\text{n} \text{U}_\text{i}$; $\text{E}_\text{i} \text{U}_\text{n}$, respectively} . \label{su1}
\end{figure}

We consider other decay processes, $\text{K}^+ \to \pi^0 \text{l}^+ \nu, \text{K} \to \text{l} \nu, \tau \to \text{K} \nu $, which give rise to the constraint on the flavor of non-universality. For simplify, we consider the ratios:  $\frac{\Gamma (\text{K} \to \mu \bar{\nu})}{\Gamma(\text{K}\to e \bar{\nu})}, \frac{\Gamma (\tau \to \text{K} \nu)}{\Gamma(\text{K}\to e \bar{\nu})},  \frac{\Gamma (\text{K}^+ \to \pi^0 \bar{\mu}\nu)}{\Gamma(\text{K}^+\to \pi^0 \bar{e} \nu)}$. In the considered model, we obtain
\bea
\frac{\Gamma (\text{K} \to \mu \bar{\nu})}{\Gamma(\text{K}\to e \bar{\nu})}&=& \frac{\sum_k |\text{C}_{2k}^{us}|^2}{\sum_k|\text{C}_{1k}^{us}|^2}\times\left[\frac{\sum_k| \text{C}^{us}_{1k}|^2}{ \sum_k|\text{C}^{us}_{2k}|^2}\right]_ {\text{SM}} \times\left[ \frac{\Gamma (\text{K} \to \mu \bar{\nu})}{\Gamma(\text{K}\to e \bar{\nu})}\right]_{\text{SM}}, \crn \nonumber
\frac{\Gamma (\tau \to \text{K} \nu)}{\Gamma(\text{K}\to e \bar{\nu})}&=& \frac{\sum_k |\text{C}_{3k}^{us}|^2}{\sum_k|\text{C}_{1k}^{us}|^2}\times\left[\frac{\sum_k| \text{C}^{us}_{1k}|^2}{ \sum_k|\text{C}^{us}_{3k}|^2}\right]_ {\text{SM}}  \times\left[ \frac{\Gamma (\tau \to \text{K} \nu)}{\Gamma(\text{K}\to e \bar{\nu})}\right]_{\text{SM}}, \crn \nonumber
\frac{\Gamma (\text{K}^+ \to \pi^0 \bar{\mu}\nu)}{\Gamma(\text{K}^+\to \pi^0 \bar{e} \nu)} &=&
\frac{\sum_k |\text{C}_{2k}^{us}|^2}{\sum_k|\text{C}_{1k}^{us}|^2}\times\left[\frac{\sum_k| \text{C}^{us}_{1k}|^2}{ \sum_k|\text{C}^{us}_{2k}|^2}\right]_ {\text{SM}}  \times \left[ \frac{\Gamma (\text{K}^+ \to \pi^0 \bar{\mu}\nu)}{\Gamma(\text{K}^+\to \pi^0 \bar{e} \nu)}\right]_{\text{SM}}.
\eea
The experimental values for these ratios are given in \cite{Workman:2022ynf}
\bea
\left[\frac{\Gamma (\text{K} \to \mu \bar{\nu})}{\Gamma(\text{K}\to e \bar{\nu})} \right]_{\text{exp}}=4.018(3) \times 10^4, \left[\frac{\Gamma (\tau \to \text{K} \nu)}{\Gamma(\text{K}\to e \bar{\nu})} \right]_{\text{exp}}=1.89(3)\times 10^7, 
\left[\frac{\Gamma (\text{K}^+ \to \pi^0 \bar{\mu}\nu)}{\Gamma(\text{K}^+\to \pi^0 \bar{e} \nu)} \right]_{\text{exp}}
 =0.660(3),
\crn \eea
while the SM predicted values can be found in\cite{Workman:2022ynf} 
\bea
\left[\frac{\Gamma (\text{K} \to \mu \bar{\nu})}{\Gamma(\text{K}\to e \bar{\nu})} \right]_{\text{SM}}=4.0037(2) \times 10^4, \left[\frac{\Gamma (\tau \to \text{K} \nu)}{\Gamma(\text{K}\to e \bar{\nu})} \right]_{\text{SM}}=1.939(4)\times 10^7, 
\left[\frac{\Gamma (\text{K}^+ \to \pi^0 \bar{\mu}\nu)}{\Gamma(\text{K}^+\to \pi^0 \bar{e} \nu)} \right]_{\text{SM}}
=0.663(2).\crn
\eea
Based on the constraints given in the previous studies, we continue our numerical analysis of the s-u transition processes. In  Figs.(\ref{su1}), we create a contour of the ratios 
$\frac{\Gamma (\text{K} \to \mu \bar{\nu})}{\Gamma(\text{K}\to e \bar{\nu})}, \frac{\Gamma (\tau \to \text{K} \nu)}{\Gamma(\text{K}\to e \bar{\nu})}$ and $\frac{\Gamma (\text{K}^+ \to \pi^0 \bar{\mu}\nu)}{\Gamma(\text{K}^+\to \pi^0 \bar{e} \nu)}$, in plane $\delta m-m_{U_1}$. The frames from top to bottom are considered according to the following cases:
$m_{E_1}=m_{E_2}=m_{E_3}, m_{U_1}=m_{U_2}=m_{U_3}$; $\text{E}_\text{n} \text{U}_\text{i}$; $\text{E}_\text{i} \text{U}_\text{n}$. In all three cases, the allowed parameter space region of $\delta m$ that can explain these experimental values, is also subdivided into the electroweak scale or the TeV scale. The allowed parameter space regions are determined by their consistency with the experimental values of $\text{R}(\text{D}), \text{R}(\text{D}^{(*)})$, and $\text{R}(\text{X}_c)$ as previously considered.    

\subsection{$d \rightarrow u $ transition}
\begin{figure}[H]
	\centering
	\includegraphics[width = 1\textwidth]{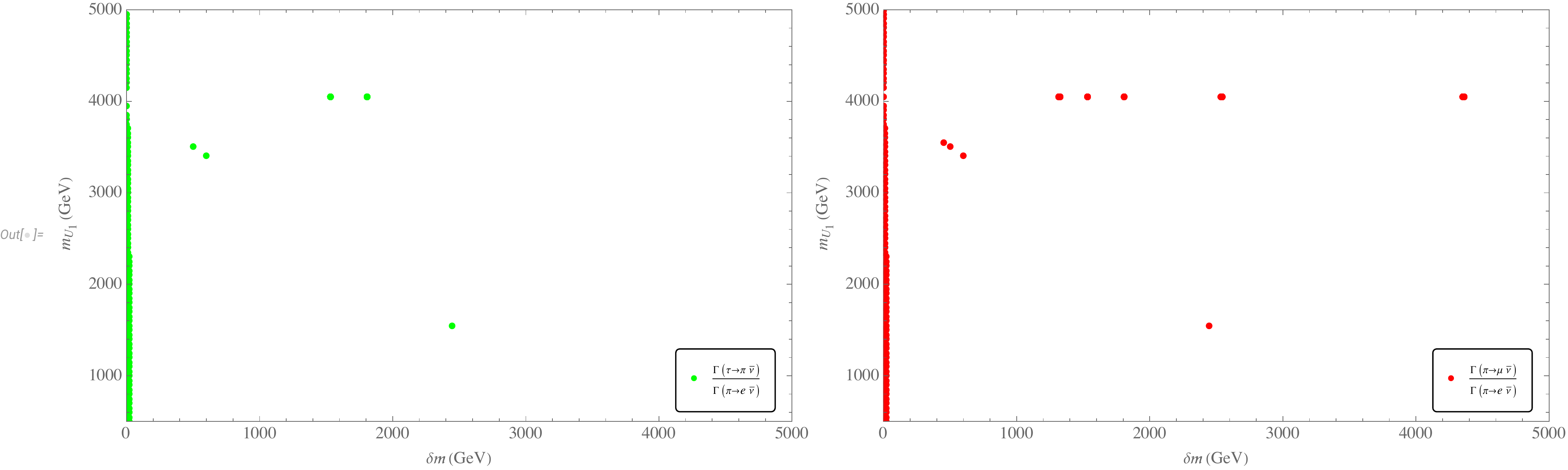}
	\includegraphics[width = 1\textwidth]{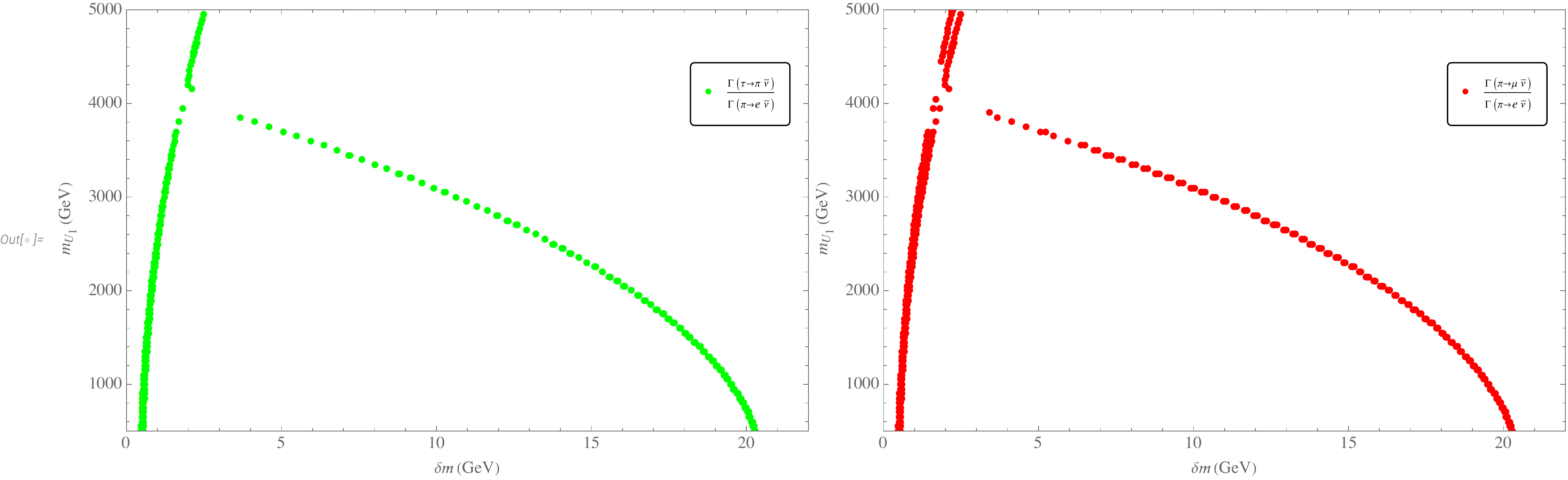}
	\caption{The ratios, $\frac{\Gamma(\tau\rightarrow\pi \nu )}{\Gamma(\pi\rightarrow e \bar{\nu})} $, $\frac{\Gamma(\pi\rightarrow\mu\bar{\nu})}{\Gamma(\pi\rightarrow e \bar{\nu})}$ corresponding to left to right frame, contoured as a
		function of $\delta m, m_{U_1}$ in the case: $m_{E_1}=m_{E_2}=m_{E_3}$, $m_{U_1}=m_{U_2}=m_{U_3}$. }
	\label{du1}
\end{figure}

\begin{figure}[H]
	\centering
	\includegraphics[width = 1\textwidth]{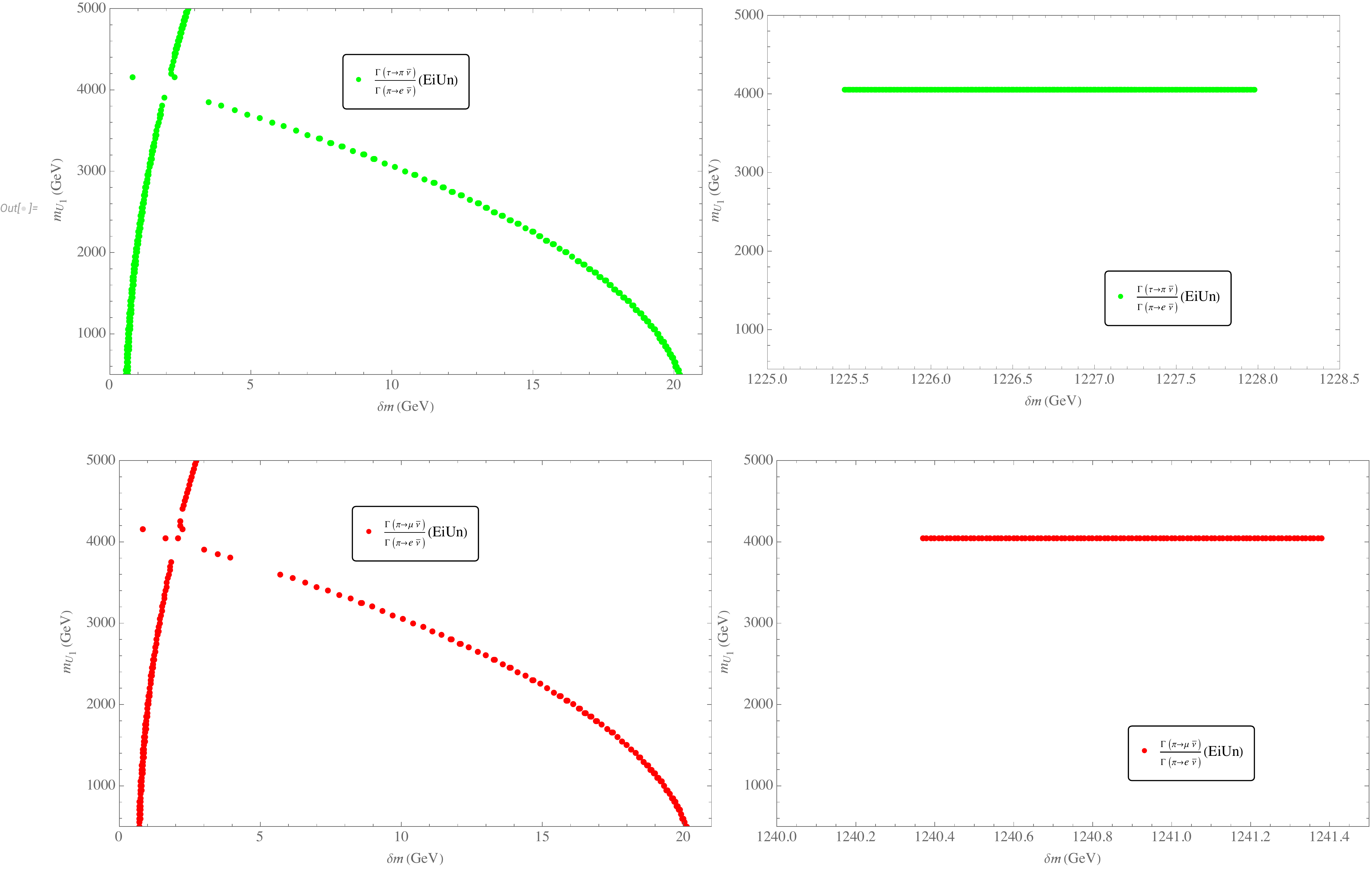}
	\includegraphics[width = 1\textwidth]{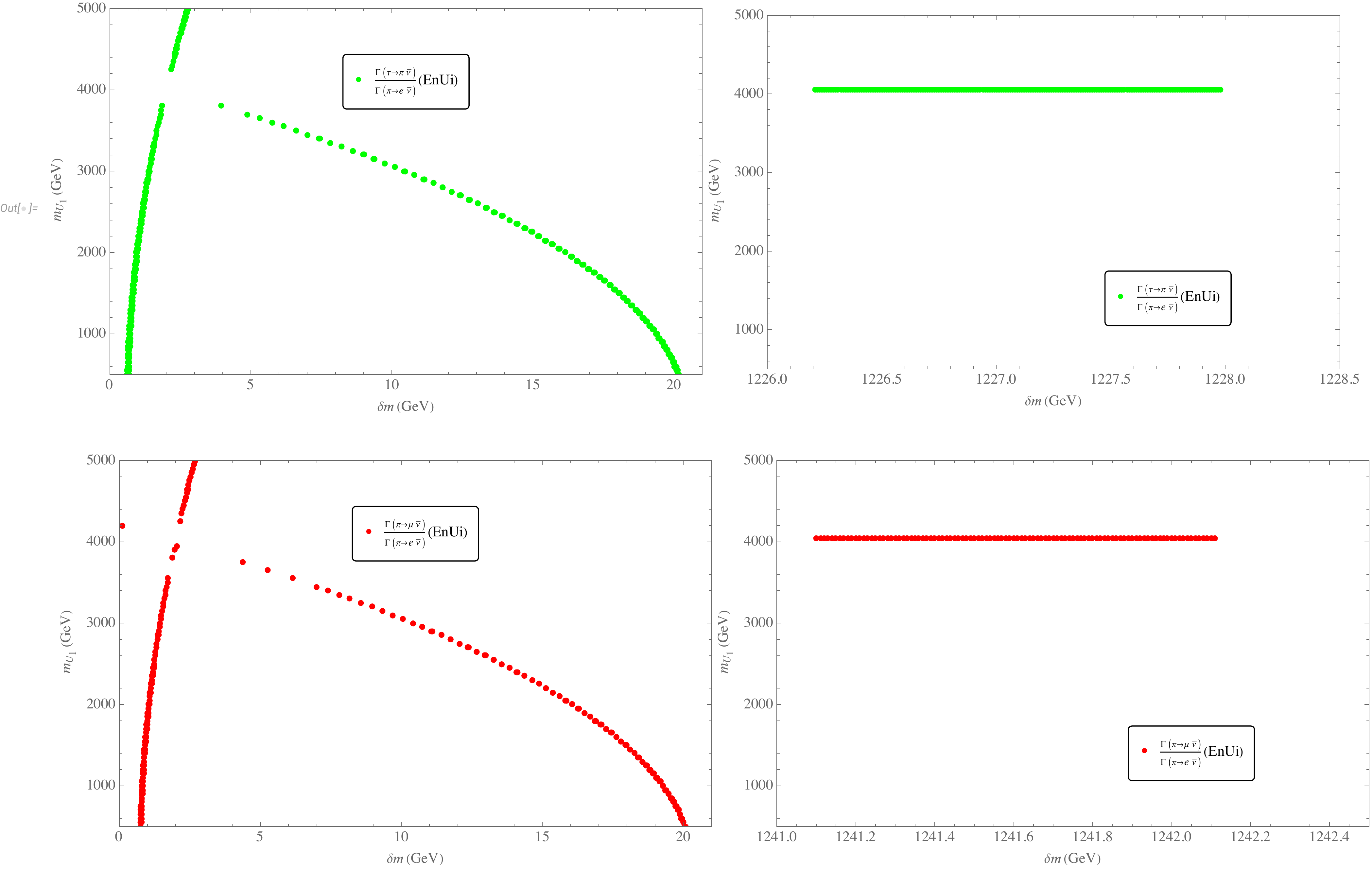}
	\caption{The ratios, $\frac{\Gamma(\tau\rightarrow\pi \nu )}{\Gamma(\pi\rightarrow e \bar{\nu})} $,  $\frac{\Gamma(\pi\rightarrow\mu\bar{\nu})}{\Gamma(\pi\rightarrow e \bar{\nu})}$, are contoured as a function of $\delta m, m_{U_1}$ in two cases: $\text{E}_{\text{i}}\text{U}_{\text{n}}$,$\text{E}_{\text{n}}\text{U}_{\text{i}}$. }
	\label{du2}
\end{figure}

One of the tighter constraints on flavor non-universality is the decay processes $d \to u l \bar{\nu}$, which corresponds to
$\pi \to l \bar{\nu}$. To cancel the dependence the combination $ \text{G}_\text{F}\mid \text{V}_{\text{ud}}\mid$,  we consider the ratios $\frac{\Gamma(\tau\rightarrow\pi\nu)}{\Gamma(\pi \rightarrow e \bar{\nu})}$ and  $\frac{\Gamma(\pi\rightarrow\mu\bar{\nu})}{\Gamma(\pi\rightarrow e \bar{\nu})}$. The experimental values for  
these ratios are collected by \cite{Boucenna:2016qad} 
\bea
\left[\frac{\Gamma(\tau\rightarrow\pi\nu)}{\Gamma(\pi \rightarrow e\bar{\nu})}\right]_{\text{exp}}= 7.90(5) \times 10^7, \hs \left[ \frac{\Gamma(\pi\rightarrow\mu\bar{\nu})}{\Gamma(\pi\rightarrow e \bar{\nu)}} \right]_{\text{exp}}= 8.13(3) \times 10^3,
\eea
while the predictive values of the SM are \cite{Cirigliano:2007ga, Boucenna:2016qad}
\bea
\hs
\left[\frac{\Gamma(\tau\rightarrow\pi\nu)}{\Gamma(\pi \rightarrow e \bar{\nu})}\right]_{\text{SM}}= 7.91(1) \times 10^7, \hs \left[ \frac{\Gamma(\pi\rightarrow\mu\bar{\nu})}{\Gamma(\pi\rightarrow e\bar{\nu})} \right]_{\text{SM}}=8.096(1) \times 10^3 .
\eea
For these ratios, the MF331 model predicts:
\bea
\frac{\Gamma(\pi\rightarrow\mu\bar{\nu})}{\Gamma(\pi\rightarrow e \bar{\nu})} &= &\frac{\sum_k\mid \text{C}^{ud}_{2k}\mid^2}{\sum_k\mid \text{C}^{ud}_{1k}\mid^2} \times\left[\frac{\sum_k| \text{C}^{ud}_{1k}|^2}{ \sum_k|\text{C}^{ud}_{2k}|^2}\right]_ {\text{SM}} \times \left[\frac{\Gamma(\pi\rightarrow\mu\bar{\nu})}{\Gamma(\pi\rightarrow e\bar{\nu})} \right]_\text{SM}, \crn \nonumber
\frac{\Gamma(\tau\rightarrow\pi \nu )}{\Gamma(\pi\rightarrow e \bar{\nu})} & =& \frac{\sum_k\mid \text{C}^{ud}_{3k}\mid^2}{\sum_k\mid \text{C}^{ud}_{1k}\mid^2}\times\left[\frac{\sum_k| \text{C}^{ud}_{1k}|^2}{ \sum_k|\text{C}^{ud}_{3k}|^2}\right]_ {\text{SM}} \times \left[\frac{\Gamma(\tau \rightarrow\mu\bar{\nu})}{\Gamma(\pi\rightarrow e\bar{\nu})} \right]_\text{SM}.
\eea
In Figs. (\ref{du1}), (\ref{du2}), we contour the ratios, $\frac{\Gamma(\tau\rightarrow\pi \nu )}{\Gamma(\pi\rightarrow e \bar{\nu})} $,  $\frac{\Gamma(\pi\rightarrow\mu\bar{\nu})}{\Gamma(\pi\rightarrow e \bar{\nu})}$, as a function of $m_{U_1}, \delta m$ in the possible cases indicated in two above sections. For the case, $m_{E_1}=m_{E_2}=m_{E_3}, m_{U_1}=m_{U_2}=m_{U_3}$, we realize that 
in the TeV scale region, there are a few values of $\delta m$ that predict the ratios, $\frac{\Gamma(\tau\rightarrow\pi \nu )}{\Gamma(\pi\rightarrow e \bar{\nu})} $,  $\frac{\Gamma(\pi\rightarrow\mu\bar{\nu})}{\Gamma(\pi\rightarrow e \bar{\nu})}$, consistent with the experimental values, while the GeV region of $\delta m$ is predicted for explaining these values. In the limit, $2 \text{GeV}<\delta m < 20 \text{GeV}$, the upper bound of $m_{U_1}$ is smaller than 4 TeV. These conclusions also apply to cases: $\text{E}_\text{i}\text{U}_\text{n}, \text{E}_\text{n}\text{U}_\text{i}$. 
\newpage
Let us review the allowed parameter space derived from studying the transition processes, $b-c,s-u,d-u$. The region of parameter space in the first scenario, where $m_{E_1}=m_{E_2}=m_{E_3}, m_{U_1}=m_{U_2}=m_{U_3}$, is determined by the intersection of planes $m_{U_1}-\delta m$ shown in Figs. (\ref{RDDX1}),(\ref{su1}), and (\ref{du1}). We get to the conclusion that the allowed region is a part of the plane limited by  $\delta m, m_{U_1}$ is: $2 <\delta m < 20$ GeV, and $m_{U_1} <  4 $ TeV or $\delta m < 2$ TeV. The parameter space derived from the scenarios, $\text{E}_{\text{i}}\text{U}_\text{n}$ ,$\text{E}_\text{n}\text{U}_\text{i}$, 
must be simultaneously consistent with the values depicted in Figs. (\ref{RDDX2}),(\ref{du2}).
Specifically, in the GeV energy scale, there is no common pair of values, $m_{U_1}-\delta m$, while in the TeV energy scale, there is a narrow region of $m_{U_1}-\delta m$ that allows to successfully accommodate the experimental values of the lepton flavor universality observables.  

 \subsection{Revising for $\text{R}_\text{K}$, $\text{R}_{\text{K}^*}$} 
\begin{figure}[H]
 	\centering
 	\includegraphics[width = 0.60\textwidth]{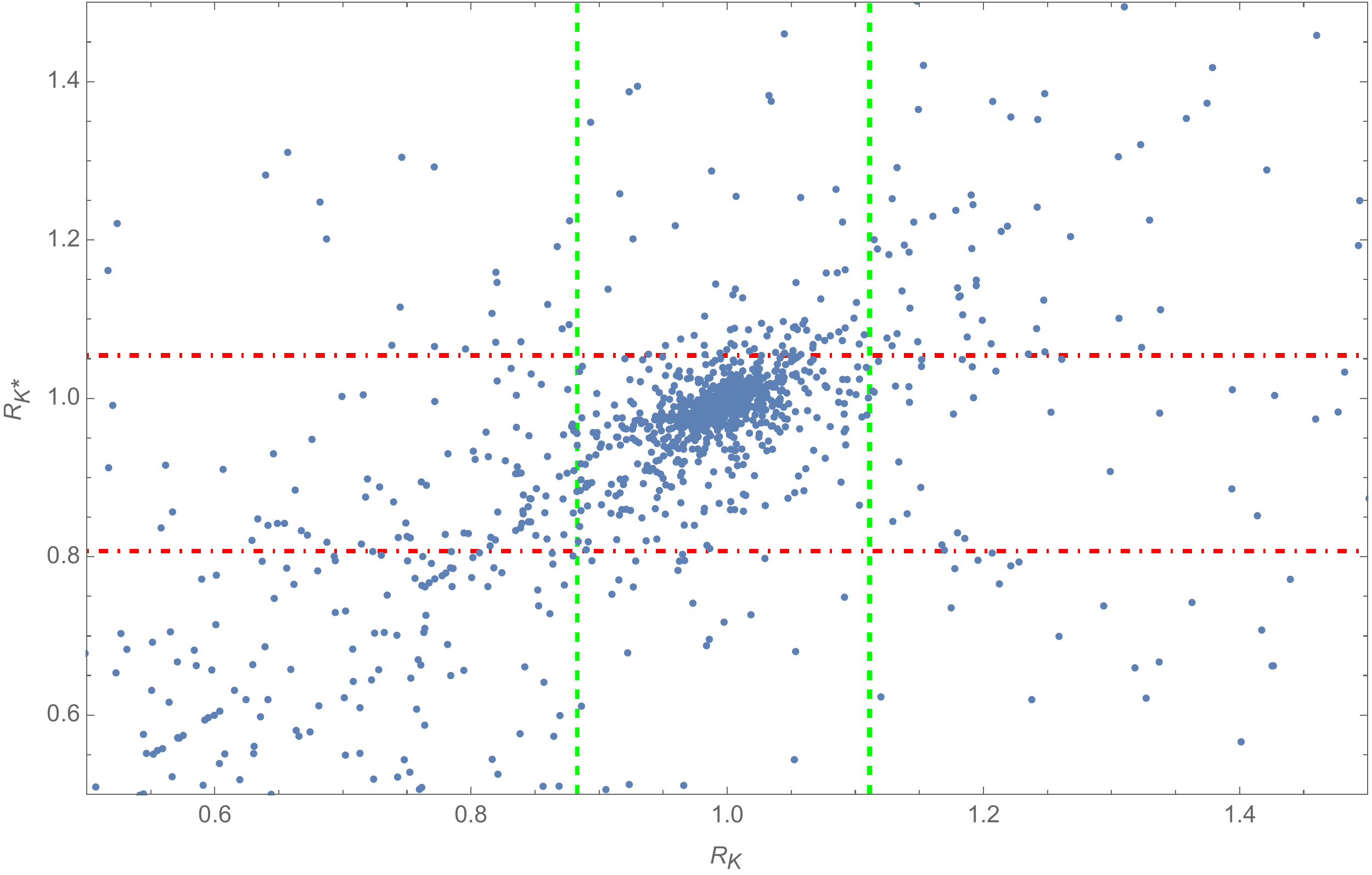}
 	\includegraphics[width = 0.60\textwidth]{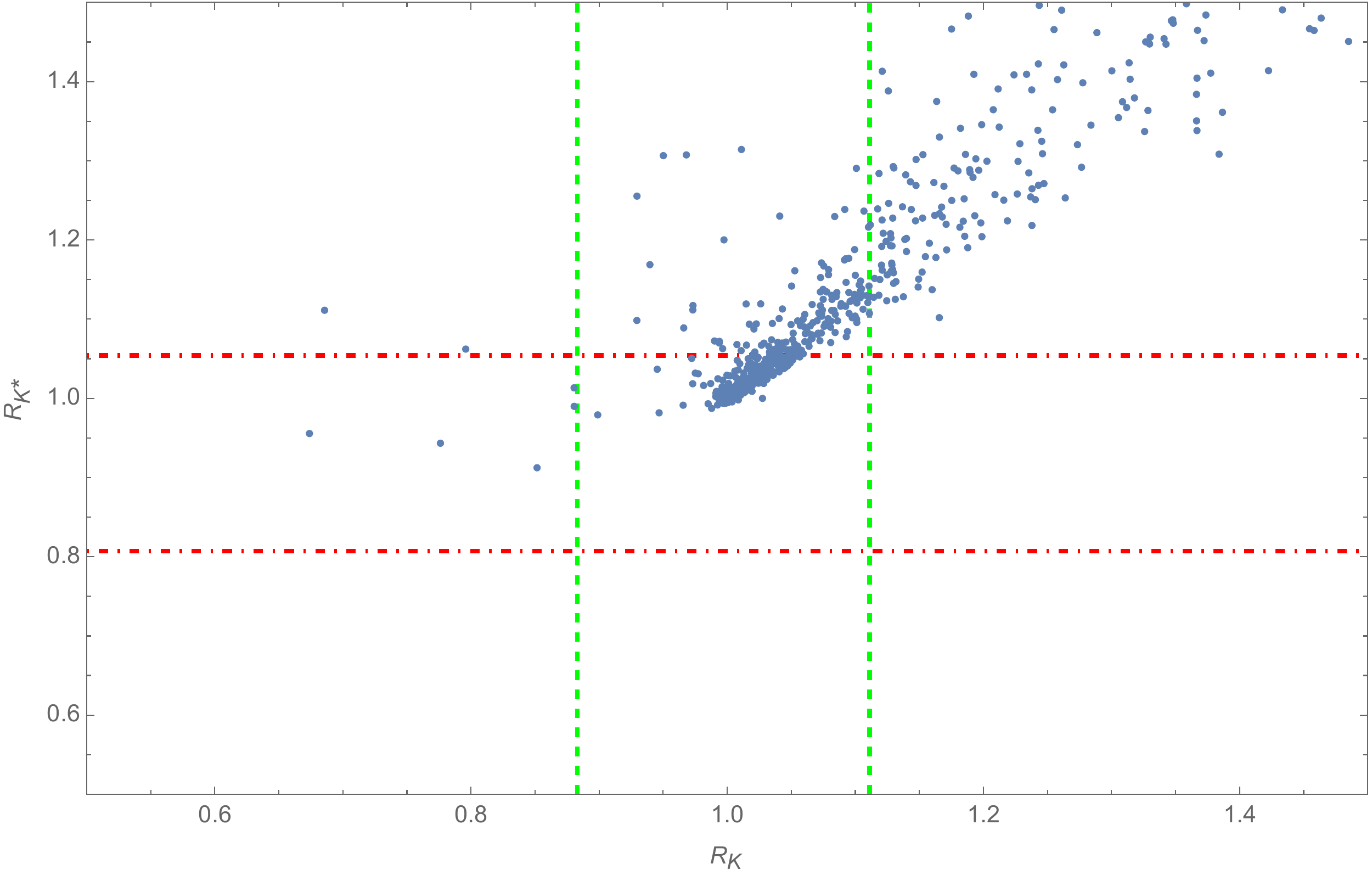}
 	\includegraphics[width = 0.60\textwidth]{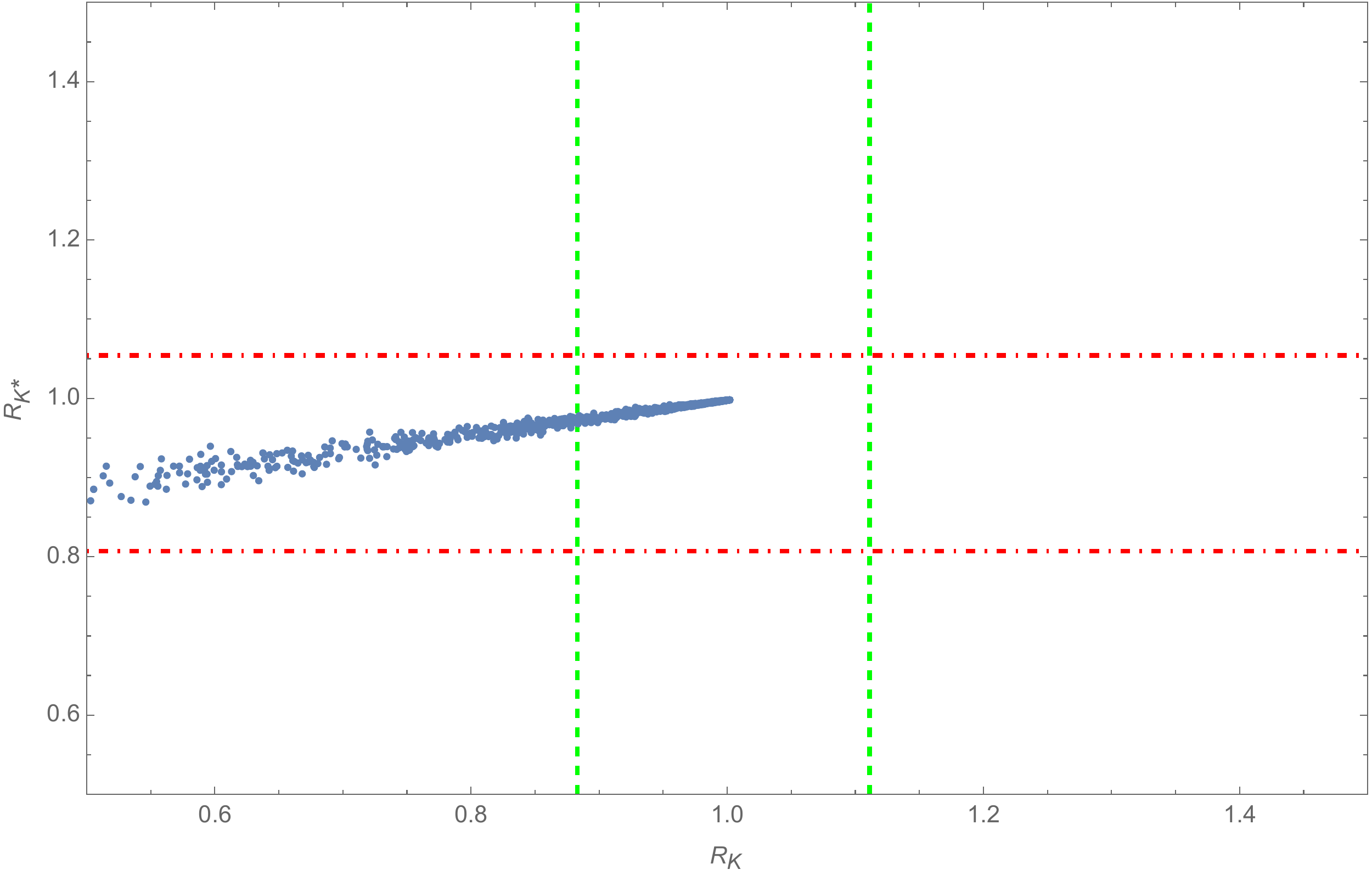}
 	\caption{The correlation between $\text{R}_\text{K}$ and $\text{R}_{\text{K}^*}$ by randomly seeding the values of $\delta m, m_{U_1}$ in the domain: $\delta \in [2, 20]$ GeV, $m_{U_1} \in [200, 5000]$ GeV. From bottom to top, the scenarios are $m_{E_1}=m_{E_2}=m_{E_3},m_{U_1}=m_{U_2}=m_{U_3}$; $\text{E}_{\text{i}}\text{U}_{\text{n}}$; $\text{E}_{\text{n}}\text{U}_{\text{i}}$. The green (red) lines represent the upper and lower limits of the most recent $\text{R}_\text{K}$($\text{R}_{\text{K}^*}$) measurements \cite{lhcbcollaboration2022test,lhcbcollaboration2022measurement} }
 	\label{RKRKs}
 \end{figure}

 In \cite{Duy:2022qhy}, we investigated the ratios $\text{R}_\text{K}$, $\text{R}_{\text{K}^*}$, corresponding to lepton flavor universality observables, in the MF331 model. We demonstrated that there are two sources of contributions to the ratios $\text{R}_\text{K}$, $\text{R}_{\text{K}^*}$, namely box, and penguin diagrams, with the box diagram having a stronger influence. With some other assumptions attached, we showed that only the box new particles' mass degeneracy can account for fitting the $b \to s \mu^+ \mu^-$ anomaly data \cite{Aaij_2014, Aaij_2019, Choudhury_2021, Aaij2021, Aaij_2017, Wehle_2021, Aaij2021}.
 December 2022, an updated LHCb analysis of $\text{R}_\text{K}$, $\text{R}_{\text{K}^*}$ based on the full Run 1 and 2 datasets has been presented \cite{lhcbcollaboration2022test,lhcbcollaboration2022measurement}. These new results are consistent with the SM predictions. These results dramatically change the scenario of NP effects in the ratios $\text{R}_\text{K}$, $\text{R}_{\text{K}^*}$. So, in the considered model, we provide a reassessment of NP effects in the $\text{R}_\text{K}, \text{R}_{\text{K}^*}$. Fig.(\ref{RKRKs}) displays the predicted outcomes, the ratios $\text{R}_\text{K}$, $\text{R}_{\text{K}^*}$, which are obtained by taking random values for $\delta m \in [2, 20]$ GeV, $m_{U_1} \in [200, 5000]$ GeV in three scenarios: $m_{E_1}=m_{E_2}=m_{E_3},m_{U_1}=m_{U_2}=m_{U_3}$, and $\text{E}_{\text{i}}\text{U}_{\text{n}}$, $\text{E}_{\text{n}}\text{U}_{\text{i}}$. We observe that the distribution of points demonstrating the correlation between $\text{R}_\text{K}$ and $\text{R}_{\text{K}^*}$depends on the mass hierarchy of new quarks (leptons).  The larger distribution density responds to the most recent measurements  \cite{lhcbcollaboration2022test,lhcbcollaboration2022measurement}, $\text{R}_\text{K}$, $\text{R}_{\text{K}^*}$, corresponds to the  
 scenarios, $m_{E_1}=m_{E_2}=m_{E_3},m_{U_1}=m_{U_2}=m_{U_3}$, and $ \text{E}_{\text{i}}\text{U}_{\text{n}}$. The model also predicts pairs of $\text{R}_\text{K}$, $\text{R}_{\text{K}^*}$ values that are consistent with the results before December 2022 \cite{Aaij_2014,Aaij_2019,Choudhury_2021,Aaij2021,Aaij_2017,Wehle_2021, Aaij2021} in the case of $m_{E_1}=m_{E_2}=m_{E_3},m_{U_1}=m_{U_2}=m_{U_3}$, but the density of matches is lower. This is not a conclusion in the case  $\text{E}_{\text{i}}\text{U}_{\text{n}}$. Compared with the two mentioned cases, the distribution of points in the case, $\text{E}_{\text{n}}\text{U}_{\text{i}}$, is completely different. The correlation between the ratios,$\text{R}_\text{K}$, $\text{R}_{\text{K}^*}$, is almost linearly distributed. In this case, not only is there a parameter space that can accommodate the
 most recent measurements of the ratios $\text{R}_\text{K}$, $\text{R}_{\text{K}^*}$ but there is also another parameter space that accommodates the old data of the $\text{R}_\text{K}$, $\text{R}_{\text{K}^*}$ observables \cite{Aaij_2014, Aaij_2019, Choudhury_2021, Aaij2021, Aaij_2017, Wehle_2021, Aaij2021}. 
 
\section{Conclusion\label{Concl.}}
We have performed a detailed analysis of the $\text{b} \to \text{c} \tau \nu$ and $\text{b} \to \text{s} l^+ l^-$ processes within the framework of theory with $SU(3)_C\times SU(3)_L\times U(1)_N$ gauge symmetry ($331$ model) with non-universal $SU(3)_L \times U(1)_N$ symmetry in the lepton sector. The theory under consideration corresponds to the MF331 model, where the scalar sector is only composed of two $SU(3)_L$ scalar triplets, one responsible for the spontaneous breaking of the $SU(3)_L\times U(1)_N$ and the other for triggering the breaking of the SM gauge group. The non universal $SU(3)_L \times U(1)_N$ assignments in the lepton sector allow for non-universal neutral and charged currents involving heavy non SM gauge bosons and SM leptons. Those interactions give rise to radiative contributions to the $\text{u}_i-\text{d}_j$ transitions arising from one loop level penguin and box diagrams involving the virtual exchange of heavy non-SM up type quarks, exotic charged leptons, and non-SM gauge bosons running in the internal lines of the loop. We performed a detailed numerical analysis of several observables related to the $b\to c$, $s\to u$ and $d\to u$ transitions, finding that the mass hierarchy of exotic quarks and new leptons has an impact on the ratios involved in these transitions. The mass differential of two new leptons, $\delta m$ at the electroweak scale and the mass of exotic up-type quarks at the TeV scale, making these particles accessible at the LHC scale, is  the region of parameter space where the observables associated with these transitions are consistent with their corresponding experimental values. In these allowable parameter space regions, the ratios, $\text{R}_\text{K}$, $\text{R}_{\text{K}^*}$, are predicted to match the recent measurements.
	
\section*{Acknowledgments}\vspace{-0.4cm}
This research is funded by International Centre of Physics, Institute of Physics, Vietnam Academy of Science and Technology under Grant No. ICP-2023.02. AECH has received funding from ANID-Chile FONDECYT 1210378, ANID-Chile PIA/APOYO AFB220004 and Milenio-ANID-ICN2019\_044
\appendix

\section{The parameters appeared in the lepton mass matrix}\label{app1}
The expressions of the functions $f^{EE}_{a b},f^{eE}_{a b},f^{ee}_{ab}, f^{e\xi}_{1b},f^{E\xi}_{1b}$ in the mass mixing matrix of leptons $\mathcal{M}_l $ are given by

\bea
 &&f^{ee}_{\al b}(v^\prime, w^\prime, s^e_{\al b}(v,w)) = -\fr{1}{\sqrt{2}}s^e_{\al b}v^\prime ,\\
  &&f^{ee}_{1 b}(v^\prime, w^\prime, s^e_{1b}(v,w)) = - \fr{1}{\sqrt{2}}\fr{s^e_{1b}}{\Lambda}v^\prime w - \fr{1}{\sqrt{2}}\fr{s'^e_{1b}}{\Lambda}vw^\prime,\\
&&f^{EE}_{\al b}(v^\prime, w^\prime, s^E_{\al b}(v,w)) = -\fr{1}{\sqrt{2}}s^E_{\al b }w^\prime,\\
 &&f^{EE}_{1 b}(v^\prime, w^\prime, s^{\prime E}_{1b}(v,w)) = - \fr{1}{2}\fr{s^E_{1b}}{\Lambda}ww^\prime -
  \fr{1}{2}\fr{s'^E_{1b}}{\Lambda}w^{\prime2},\\
&&f^{eE}_{\al b}(v^\prime, w^\prime, s^E_{\al b}(v,w)) = -\fr{1}{\sqrt{2}}h^E_{\al b}v^\prime -\fr{1}{\sqrt{2}}s^E_{\al b}v  - \fr{1}{\sqrt{2}}h^e_{\al b}w^\prime,\\
&&f^{eE}_{1b}(v^\prime, w^\prime, s^{\prime E}_{1b}(v,w)) = -\fr{1}{\sqrt{2}}\fr{h^E_{1 b}}{\Lambda}v^\prime w  - \fr{1}{2\sqrt{2}}\fr{s^E_{1b}}{\Lambda}(v^\prime w^\prime+vw)- \fr{1}{\sqrt{2}}\fr{s^{\prime E}_{1b}}{\Lambda}v w^\prime,\\
&&f^{e\xi}_{1 b}(w^\prime,v^\prime,s_{1b}^ew,\text{s}_{1b}^ev) = -\fr {h^e_{1b}}{ 2\Lambda}vv^\prime - \frac{s^e_{1b}}{2\Lambda}v^\prime - \frac{s^{\prime e}_{1b}}{2\Lambda}v^2,\\
 &&f^{E\xi}_{1 b}(w^\prime,v^\prime,s_{1b}^ew,\text{s}_{1b}^ev) = -\fr {h^E_{1b}}{ 2\Lambda}v^{\prime 2} - \frac{s^E_{1b}}{2\Lambda}vv^\prime - \frac{s^{\prime E}_{1b}}{2\Lambda}v^2.
 \eea
 \section{The $\Ga^{f_if_jf_k}$ functions }\label{app2}
 \bea 
 &&\Ga^{WZe_b} = (m^2_Z-m^2_W)\left(1 +\fr{1}{\ep} - \ga + \ln{4\pi} -\ln m^2_{e_b}\right) - m^2_Z\frac{\ln{ x_Z^{e_b}}}{x_Z^{e_b}-1} + m^2_W\frac{\ln{ x_W^{e_b}}}{x_W^{e_b}-1},\\
 &&\Ga^{WZ\nu_a} = (m^2_Z-m^2_W)\left(1 + \fr{1}{\ep} - \ga + \ln{4\pi} -\ln m^2_{\nu_a}\right) - m^2_Z\frac{\ln{ x_Z^{\nu_a}}}{x_Z^{\nu_a}-1} + m^2_W\frac{\ln{ x_W^{\nu_a}}}{x_W^{\nu_a}-1},\\ 
&&\Ga^{W\nu_a e_b}_Z= (m^2_{e_b}-m^2_{\nu_a})\left(1 +\fr{1}{\ep} - \ga + \ln{4\pi} -\ln m^2_Z\right) - m^2_{e_b}\frac{\ln{ x_{e_b}^Z}}{x_{e_b}^Z-1} + m^2_{\nu_a}\frac{\ln{ x_{\nu_a}^Z}}{x_{\nu_a}^Z-1},\\
&&\Ga^{W\nu_a e_b}_{Z^\prime}= (m^2_{e_b}-m^2_{\nu_a})\left(1 +\fr{1}{\ep} - \ga + \ln{4\pi} -\ln m^2_{Z^\prime}\right) - m^2_{e_b}\frac{\ln{ x_{e_b}^{Z^\prime}}}{x_{e_b}^{Z^\prime}-1} + m^2_{\nu_a}\frac{\ln{ x_{\nu_a}^{Z^\prime}}}{x_{\nu_a}^{Z^\prime}-1},\\
&&\Ga^{W\ga e_c}= \left(1 +\fr{1}{\ep} - \ga + \ln{4\pi} -\ln m^2_{e_c}\right) - \frac{\ln{ x_W^{e_c}}}{x_W^{e_c}-1} ,\\
&&\Ga^{XYE_c} = (m^2_X-m^2_Y)\left(1 +\fr{1}{\ep} - \ga + \ln{4\pi} -\ln m^2_{E_c}\right) - m^2_X\frac{\ln{ x_X^{E_c}}}{x_X^{E_c}-1} + m^2_Y\frac{\ln{ x_Y^{E_c}}}{x_Y^{E_c}-1},\\
&&\Ga^{XY\xi^0} = (m^2_X-m^2_Y)\left(1 + \fr{1}{\ep} - \ga + \ln{4\pi} -\ln m^2_{\xi^0}\right) - m^2_X\frac{\ln{ x_X^{\xi^0}}}{x_X^{\xi^0}-1} + m^2_Y\frac{\ln{ x_Y^{\xi^0}}}{y_Y^{\xi^0}-1},\\
&&\Ga^{XY\xi} = (m^2_X-m^2_Y)\left(1 + \fr{1}{\ep} - \ga + \ln{4\pi} -\ln m^2_{\xi}\right) - m^2_X\frac{\ln{ x_X^{\xi}}}{x_X^{\xi} -1} + m^2_Y\frac{\ln{ x_Y^{\xi}}}{x_Y^{\xi} -1},\\
&&\Ga^{XYU_c} = (m^2_X-m^2_Y)\left(1 + \fr{1}{\ep} - \ga + \ln{4\pi} -\ln m^2_{U_c}\right) - m^2_X\frac{\ln{x_X^{U_c}}}{x^X_{U_c}-1} + m^2_Y\frac{\ln{ x_Y^{U_c}}}{x_Y^{U_a}-1},
\eea
with $ x^a_b=\fr{m^2_a}{m^2_b} $.
 \section{The  $\Ga^{f_if_j}$ functions }\label{app3}
 \bea
\Ga^{U_lE_c}&= & \left[\frac{({x^{U_l}_X})^2}{\left(x^{U_l}_X-x^{E_c}_X\right) \left(x^{U_l}_X-1 \right)} -\frac{({x^{U_l}_Y})^2}{\left(x^{U_l}_Y-x^{E_c}_Y\right) \left(x^{U_l}_Y-1 \right)}\right] \ln m_{U_l}\crn
&-&\left[\frac{({x^{E_c}_X})^2}{\left(x^{E_c}_X-x^{U_l}_X\right) \left(x^{E_c}_X-1 \right)} -\frac{({x^{E_c}_Y})^2}{\left(x^{E_c}_Y-x^{U_l}_Y\right) \left(x^{E_lc}_Y-1 \right)}\right] \ln m_{E_c} \crn  &+& \frac{x^{U_l}_X x^{E_c}_X}{\left(x^{U_l}_X-1\right)\left(x^{E_c}_X-1\right)}\ln m_{X}
-\frac{x^{U_l}_Y x^{E_c}_Y}{\left(x^{U_l}_Y-1\right)\left(x^{E_c}_Y-1\right)}\ln m_{Y}
\eea
\bea
\Ga^{U_l\xi^0}&= & \left[\frac{({x^{U_l}_X})^2}{\left(x^{U_l}_X-x^{\xi^0}_X\right) \left(x^{U_l}_X-1 \right)} -\frac{({x^{U_l}_Y})^2}{\left(x^{U_l}_Y-x^{\xi^0}_Y\right) \left(x^{U_l}_Y-1 \right)}\right] \ln m_{U_l}\crn
&-&\left[\frac{({x^{\xi^0}_X})^2}{\left(x^{\xi^0}_X-x^{U_l}_X\right) \left(x^{\xi^0}_X-1 \right)} -\frac{({x^{\xi^0}_Y})^2}{\left(x^{\xi^0}_Y-x^{U_l}_Y\right) \left(x^{\xi^0}_Y-1 \right)}\right] \ln m_{\xi^0} \crn  &+& \frac{x^{U_l}_X x^{\xi^0}_X}{\left(x^{U_l}_X-1\right)\left(x^{\xi^0}_X-1\right)}\ln m_{X}
-\frac{x^{U_l}_Y x^{\xi^0}_Y}{\left(x^{U_l}_Y-1\right)\left(x^{\xi^0}_Y-1\right)}\ln m_{Y}
\eea
\bea
\Ga^{U_l\xi}&= & \left[\frac{({x^{U_l}_X})^2}{\left(x^{U_l}_X-x^{\xi}_X\right) \left(x^{U_l}_X-1 \right)} -\frac{({x^{U_l}_Y})^2}{\left(x^{U_l}_Y-x^{\xi}_Y\right) \left(x^{U_l}_Y-1 \right)}\right] \ln m_{U_l}\crn
&-&\left[\frac{({x^{\xi}_X})^2}{\left(x^{\xi}_X-x^{U_l}_X\right) \left(x^{\xi}_X-1 \right)} -\frac{({x^{\xi}_Y})^2}{\left(x^{\xi}_Y-x^{U_l}_Y\right) \left(x^{\xi}_Y-1 \right)}\right] \ln m_{\xi} \crn  &+& \frac{x^{U_l}_X x^{\xi}_X}{\left(x^{U_l}_X-1\right)\left(x^{\xi}_X-1\right)}\ln m_{X}
-\frac{x^{U_l}_Y x^{\xi}_Y}{\left(x^{U_l}_Y-1\right)\left(x^{\xi}_Y-1\right)}\ln m_{Y}
\eea
with $x^{a}_{b}=\frac{m_{a}^2}{m^2_b}.$

 \section{New couplings of gauge bosons}\label{app4}

\begin{figure}[h]
{\includegraphics[width=15cm]{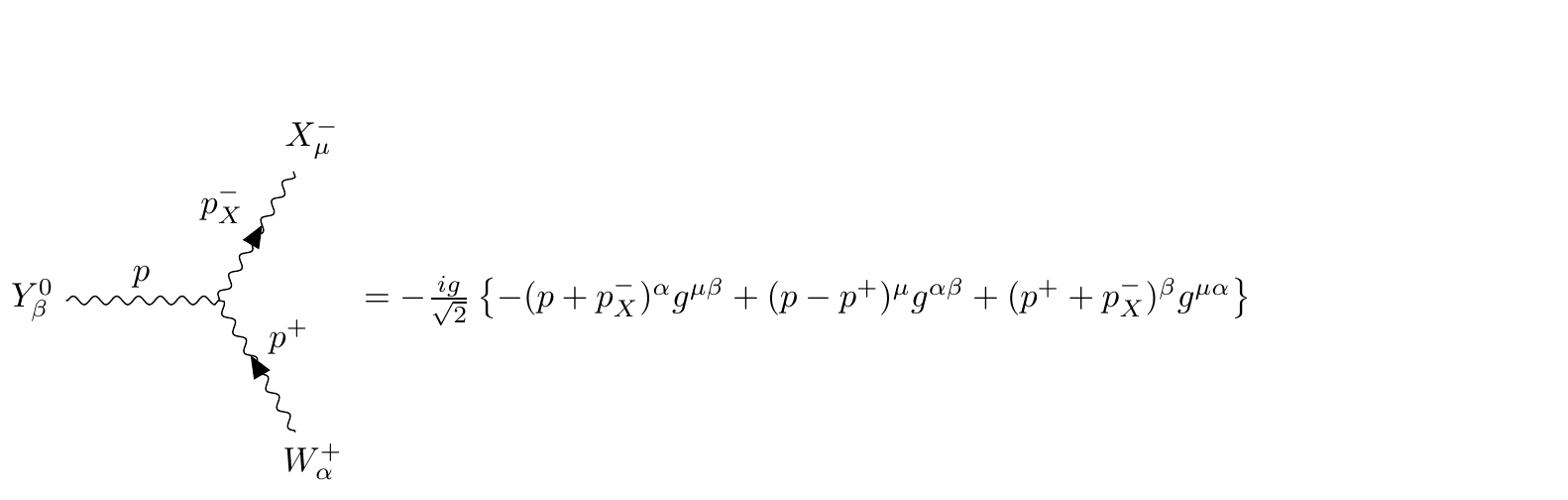}}  
\end{figure}
\begin{figure}[h]
{\includegraphics[width=15cm]{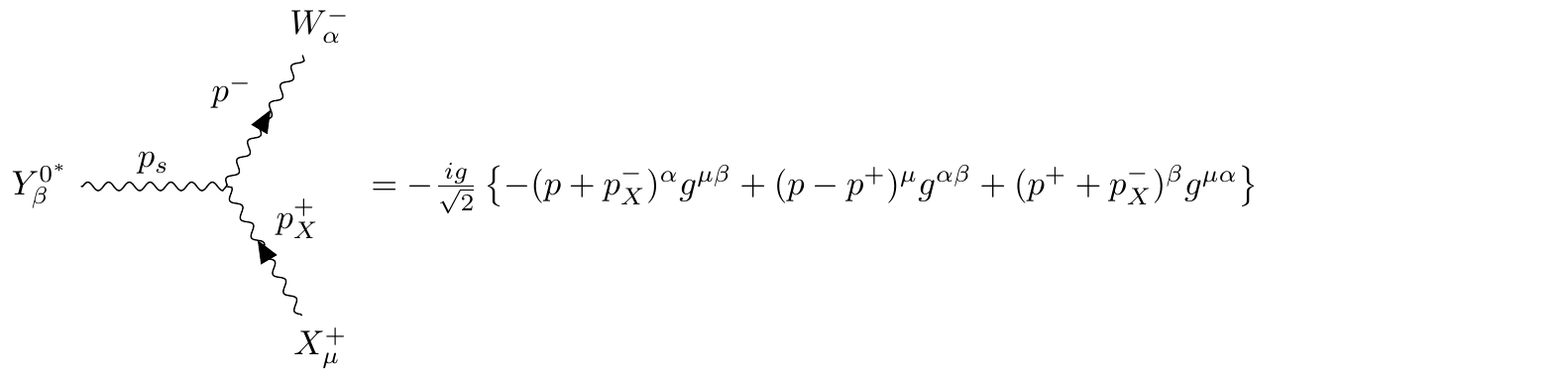}}  
\end{figure}
\begin{figure}[h]
{\includegraphics[width=15cm]{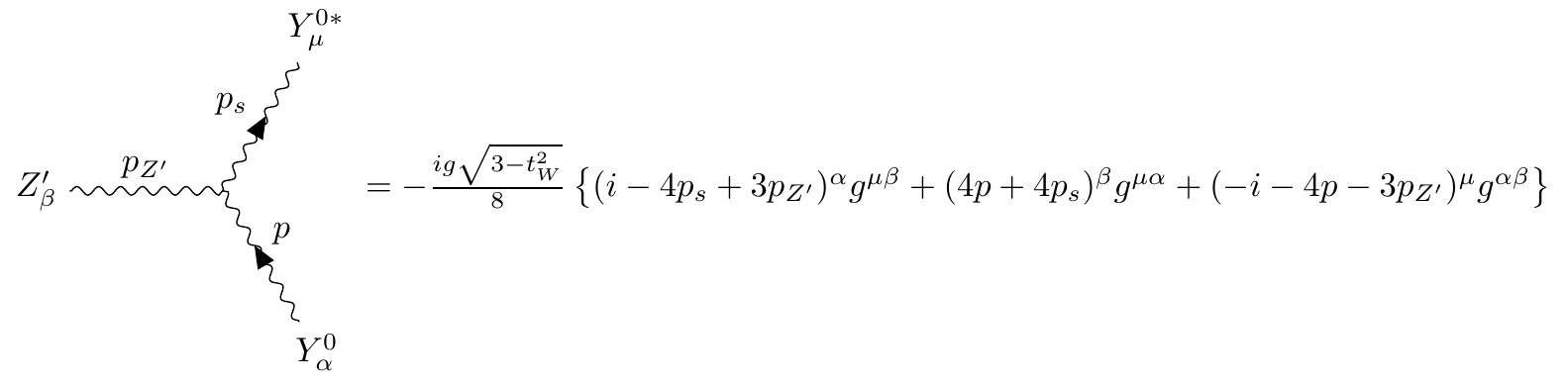}}  
\end{figure}
\begin{figure}[h]
{\includegraphics[width=15cm]{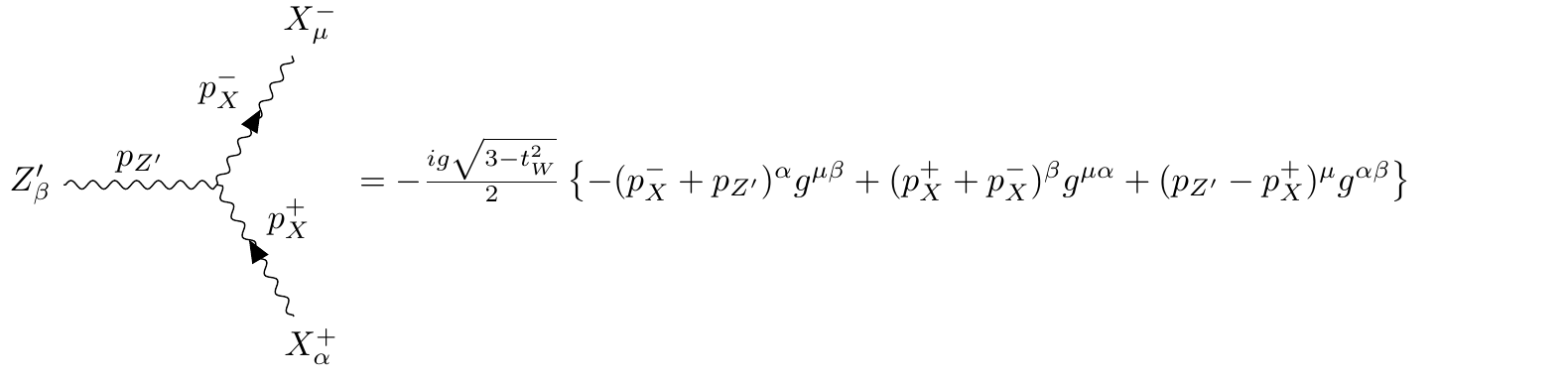}}  
\end{figure}

\newpage
\centerline{\bf{REFERENCES}}\vspace{-0.4cm} 
\bibliographystyle{utphys}
\bibliography{ThuFeMF331April3st.bib}
\end{document}